\documentclass[twocolumn,amsmath,amssymb,nofootinbib]{revtex4-2}

\usepackage{subfiles}
\usepackage{graphicx}
\usepackage{dcolumn}
\usepackage{bm}
\usepackage[
    colorlinks=true,
    citecolor=blue,
    linkcolor=blue,
    urlcolor=blue,
    pdfborder={0 0 0}
]{hyperref}
\usepackage{verbatim}
\usepackage{siunitx}
\usepackage{upgreek}
\usepackage{multirow}
\usepackage{subcaption}
\usepackage{caption}
\usepackage{ragged2e}
\graphicspath{{\subfix{Figures/}}}
\usepackage{comment}
\usepackage{rotating}
\usepackage{orcidlink}

\begin{document}

\title{\textbf{Design of the Electron-Beam Collimator for the Electron-Ion Collider}}

\author{Andrii Natochii \orcidlink{0000-0002-1076-814X}}            %
 \email{natochii@bnl.gov}
\author{Elke-Caroline Aschenauer \orcidlink{0000-0002-3856-5475}}   %
\author{Jay Best \orcidlink{0009-0005-4937-9677}}                   %
\author{Alexei Blednykh \orcidlink{0000-0001-9391-1820}}            %
\author{Xiaofeng Gu \orcidlink{0000-0002-5560-6419}}                %
\author{Kuppan Gunavathi \orcidlink{0009-0007-9527-9284}}           %
\author{Charles Hetzel \orcidlink{0009-0005-4856-9196}}             %
\author{Christoph Montag \orcidlink{0000-0003-0599-3390}}           %
\affiliation{Brookhaven National Laboratory, Upton, New York 11973, USA}

\date{\today}

\begin{abstract}
The design and optimization of a beam collimator for the Electron Storage Ring (ESR) of the Electron-Ion Collider (EIC) are presented. The collimator must efficiently intercept stray beam particles while maintaining acceptable beam coupling impedance, surviving rapid accidental beam impacts, and withstanding steady-state beam-induced heating. An integrated design approach combining beam-loss tracking, electromagnetic-shower simulations, thermomechanical analysis, and impedance calculations is developed and benchmarked against published simulations and beam-impact experiments. A comparison of candidate absorber materials identifies Al2219-T62 as providing a favorable balance of beam-impact robustness, thermal and electrical conductivity, mechanical strength, and ultra-high-vacuum compatibility. For the ultimate ESR bunch charge of \SI{28}{nC} at \SI{10}{GeV}, the Al2219-T62 jaw remains below the adopted practical thermal-shock threshold for approximately 50 fully intercepted bunches and below the localized melting threshold for approximately 100 bunches under the reference impact conditions considered. These results provide an engineering input to the machine-protection requirements, for which a complete protection response within several machine turns is required. Optimization of the absorber-tip length yields a \SI{70}{mm} tapered jaw that balances collimation performance, beam-impact robustness, and impedance. The resulting design maintains acceptable impedance margins and remains below both the yield and ultimate tensile strengths under a conservative \SI{3.6}{kW} beam-induced heating scenario. The results establish a baseline ESR collimator design and a benchmarked methodology for evaluating collimator robustness in high-current electron storage rings.
\end{abstract}

\maketitle

\section{\label{sec:Introduction}Introduction}

The Electron Storage Ring (ESR) of the Electron-Ion Collider (EIC)~\cite{osti_1765663} will operate with high-current, high-energy electron beams, imposing stringent requirements on beam-loss control and machine--detector interface (MDI) protection near the interaction region. The combination of substantial stored beam energy, continuous halo losses, and the possibility of rapidly developing beam instabilities presents significant challenges for both detector-background mitigation and machine protection. Uncontrolled beam losses in the vicinity of the interaction region can generate excessive detector backgrounds and radiation or heat loads in sensitive accelerator components and, in severe cases, may cause irreversible damage to vacuum-system and beam-intercepting components.

To mitigate these risks, the ESR collimation system must satisfy several competing requirements. The collimators must efficiently intercept stray beam-halo particles before they reach the interaction region while maintaining acceptable beam-coupling impedance and sufficient robustness against designated accidental beam impacts. In electron storage rings, electromagnetic-shower development and beam-induced heat loads place particularly stringent constraints on the choice of collimator material and geometry. These design parameters directly determine the effectiveness of downstream loss suppression, the spatial distribution and peak density of deposited energy, the resulting thermomechanical response, and the contribution of the collimators to the machine impedance.

The optimization of the ESR collimators must account for a wide range of beam-loss conditions. During routine operation, the collimators are expected to intercept continuous, low-level halo losses without compromising beam stability or lifetime. In contrast, abnormal conditions such as rapid orbit excursions or fast instability-driven beam motion can produce highly localized losses over only a few machine turns. In the most severe cases, a substantial fraction of the stored beam can be intercepted within a single revolution ($\sim\SI{12.8}{\micro\second}$), producing energy deposition sufficient to approach or exceed material-damage thresholds. The resulting turn-scale evolution of the beam loss therefore places stringent requirements on both collimator robustness and the response time of the machine protection system (MPS).

Accurate assessment of the ESR collimator performance requires an integrated treatment of beam dynamics, particle--matter interactions, thermal response, and beam coupling impedance. Multi-turn tracking and Monte Carlo particle-transport simulations are used to determine the spatial distribution of beam losses, electromagnetic-shower development, and energy deposition, and to evaluate the dependence of downstream loss suppression on the absorber material and geometry. The resulting energy-deposition distributions are used to assess the transient temperature rise and beam-impact survivability of candidate materials, while finite-element thermomechanical calculations are employed to evaluate the response to steady-state beam-induced heating. In parallel, electromagnetic wakefield calculations quantify the impedance contribution of the collimator and its compatibility with ESR beam-stability requirements. Together, these analyses provide the basis for balancing collimation efficiency, impact tolerance, thermal performance, and impedance within a single engineering design.

The remainder of this paper is organized as follows. Section~\ref{sec:collimation_overview} summarizes the ESR collimation system, relevant beam parameters, and baseline optics, while Sec.~\ref{sec:design_constraints} introduces the principal design constraints and optimization strategy. Section~\ref{sec:length_optimization} examines the dependence of downstream interaction region losses on absorber material and length. Section~\ref{sec:survivability} develops the energy-deposition and temperature-rise methodology, benchmarks it against published simulations and beam-impact studies, and applies it to the survivability assessment of candidate ESR absorber materials. The resulting operational configuration and its connection to MPS requirements are discussed in Sec.~\ref{sec:machine_protection}. Section~\ref{sec:head_refinement} then refines the tapered-jaw geometry using the combined beam-loss and survivability results. The impedance contribution and associated single-bunch stability constraints are evaluated in Sec.~\ref{sec:impedance}, followed by the steady-state thermomechanical analysis of the baseline jaw design in Sec.~\ref{sec:steady_state}. Finally, Sec.~\ref{sec:conclusions} summarizes the resulting baseline ESR collimator design and the principal conclusions of the study.

\section{\label{sec:collimation_overview}ESR Collimation System Overview}

The ESR collimation system is designed to suppress beam-induced detector backgrounds, protect superconducting (SC) magnets and other sensitive accelerator components from excessive radiation and heat loads, and localize beam losses within dedicated regions of the ring. The overall collimation concept, including the selection of collimator locations, optics optimization, and multi-turn beam-loss tracking, was established in Ref.~\cite{2v68-np87}. The present work builds on that system-level design and addresses the detailed optimization of the collimator absorber material, jaw geometry, beam-impact survivability, and impedance characteristics.

\begin{figure}[htbp]
\centering
    \includegraphics[width=\linewidth]{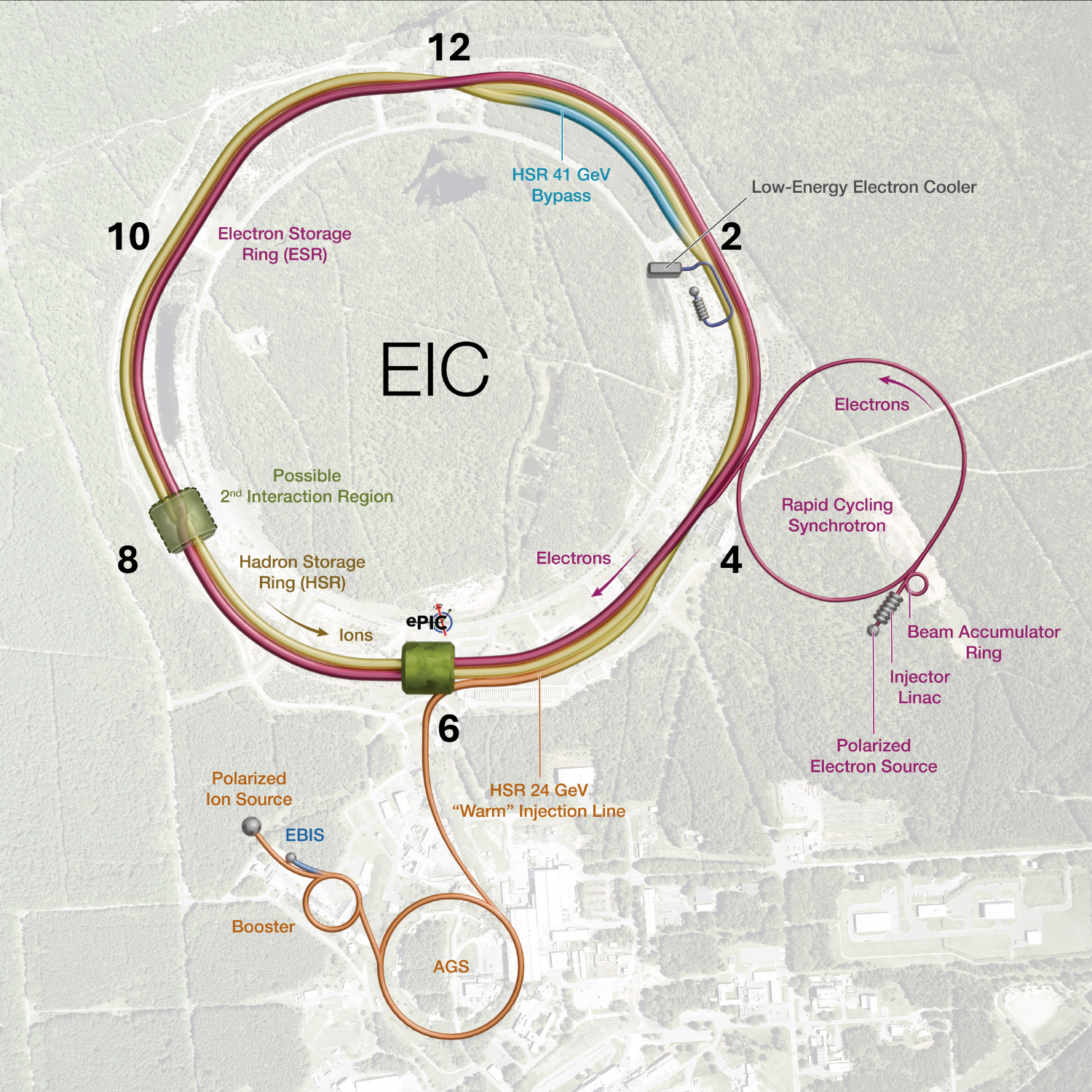}
    \caption{\label{fig:EIC_schematic_drawing}
    Schematic layout of the EIC. The numbers 2, 4, 6, 8, 10, and 12 denote the six insertion regions around the collider rings.}
\end{figure}

\begin{table*}[htbp]
\centering
    \caption{\label{tab:MachineParameters}
    Ultimate EIC beam parameters relevant to the present study, adapted from Table~3.3 of Ref.~\cite{osti_1765663}. The quantities $\varepsilon$, $\beta^{*}$, and $\Delta p/p$ denote the RMS emittance, the betatron function at the interaction point, and the RMS fractional momentum spread, respectively.}
    \begin{tabular}{l|cc|cc|cc}
    \hline\hline
    Species & proton & electron & proton & electron & proton & electron \\
    Energy [GeV] & 275 & 18 & 275 & 10 & 100 & 5 \\
    \hline
    Bunch intensity [$10^{10}$] & 19.1 & 6.2 & 6.9 & 17.2 & 4.8 & 17.2 \\
    Number of bunches & \multicolumn{2}{c|}{290} & \multicolumn{2}{c|}{1160} & \multicolumn{2}{c}{1160} \\
    Beam current [A] & 0.69 & 0.227 & 1 & 2.5 & 1 & 2.5 \\
    $\varepsilon$, H/V [nm] & 18/1.6 & 24/2 & 11.3/1 & 20/1.3 & 26/2.3 & 20/1.8 \\
    RMS bunch length [cm] & 6 & 0.9 & 6 & 0.7 & 7 & 0.7 \\
    RMS $\Delta p/p$ [$10^{-4}$] & 6.8 & 10.9 & 6.8 & 5.8 & 9.7 & 6.8 \\
    $\beta^{*}$, H/V [cm] & 80/7.1 & 59/5.7 & 80/7.2 & 45/5.6 & 61/5.5 & 78/7.1 \\
    Luminosity [$10^{33}\,\mathrm{cm^{-2}\,s^{-1}}$] &
    \multicolumn{2}{c|}{1.54} &
    \multicolumn{2}{c|}{10} &
    \multicolumn{2}{c}{3.68} \\
    \hline\hline
    \end{tabular}
\end{table*}

Figure~\ref{fig:EIC_schematic_drawing} shows the EIC ring layout, while Table~\ref{tab:MachineParameters} summarizes the beam parameters relevant to the present studies. The ESR is designed for electron beam energies of 5, 10, and \SI{18}{GeV}, with beam currents reaching \SI{2.5}{A} for the 5 and \SI{10}{GeV} configurations. Operation of such high-current beams is accompanied by intense synchrotron radiation (SR), with a total radiated power reaching approximately \SI{10}{MW}, restricted apertures in the interaction-region final-focus system of approximately $15$--$30\,\sigma_{\beta}$, and stringent detector-background requirements. Beam losses generated by Touschek scattering, beam--gas interactions, Compton scattering on thermal photons, and Bethe--Heitler processes at the interaction point can produce significant radiation loads in insertion region~6 (IR6, see Fig.~\ref{fig:EIC_schematic_drawing}), which contains the electron--Proton/Ion Collider (ePIC) detector~\cite{ePIC_web-page} and the SC final-focus magnets, as discussed in Ref.~\cite{osti_3021706}. The ESR collimation system is therefore designed to intercept the corresponding beam halo upstream of IR6 and reduce losses at the limiting apertures surrounding the detector.

The primary ESR betatron-collimation insertion is located in IR4, approximately \SI{600}{m} upstream of IR6 along the direction of beam propagation. This location provides favorable lattice optics for halo interception, sufficient longitudinal space for movable collimator assemblies, and substantial separation from the experimental region, allowing secondary particles generated at the collimators to be intercepted or lost before reaching IR6. Several horizontal and vertical collimator locations are incorporated into dedicated drift sections of the IR4 lattice, as described in detail in Ref.~\cite{2v68-np87}.

\begin{table}[htbp]
    \caption{\label{tab:Collimator_Optics_Parameters}
    Optics parameters and nominal settings of the primary ESR collimators for the three electron-beam energies. Here, $S$ denotes the distance from IP6, $\beta$ is the betatron function, $\sigma_{\beta}=\sqrt{\beta\varepsilon}$ is the RMS transverse beam size associated with betatron motion, $\varepsilon$ is the RMS emittance, $D$ is the dispersion, and $\Delta\mu$ is the betatron phase advance relative to the location of the narrowest aperture in the IR6 final-focus quadrupoles. The collimator half-aperture and the corresponding limiting IR6 aperture are denoted by $A_{\mathrm{coll.}}$ and $A_{\mathrm{IR6}}$, respectively. Quantities are given in the transverse plane corresponding to each horizontal or vertical collimator.}
    \centering
    \begin{tabular}{c|c|c|c|c|c|c|c}
        \hline\hline
        Energy & Collimator & $S$ & $\beta$ & $D$ & $\Delta\mu$ & $A_\mathrm{coll.}$ & $A_\mathrm{IR6}$ \\
        \multicolumn{1}{c|}{[GeV]} &
        \multicolumn{1}{c|}{Name} &
        \multicolumn{1}{c|}{[m]} &
        \multicolumn{1}{c|}{[m]} &
        \multicolumn{1}{c|}{[cm]} &
        \multicolumn{1}{c|}{[$2\pi$]} &
        \multicolumn{1}{c|}{[$\sigma_{\beta}$] ([mm])} &
        \multicolumn{1}{c}{[$\sigma_{\beta}$]}\\
        \hline
        \multirow{2}{*}{5}  & D04H2 & 535.53 & 90.84 & -4.00 & 0.51 & 15 (20.2) & 19\\
                             & D04V2 & 516.63 & 82.20 & 0.01  & 0.44 & 21 (8.1) & 27\\
        \hline
        \multirow{2}{*}{10} & D04H1 & 569.33 & 83.41 & 10.31 & 0.01 & 9 (11.6) & 15\\
                             & D04V1 & 584.83 & 88.99 & -0.02 & 0.04 & 21 (7.1) & 28\\
        \hline
        \multirow{2}{*}{18} & D04H3 & 524.43 & 78.19 & 3.31  & 0.48 & 12 (16.4) & 15\\
                             & D04V2 & 516.63 & 58.03 & 0.00  & 0.46 & 18 (6.1) & 23\\
        \hline\hline
    \end{tabular}
\end{table}

Table~\ref{tab:Collimator_Optics_Parameters} summarizes the nominal primary-collimator settings and local optics for each ESR beam-energy configuration. The collimation optics are based on lattice version~6.3.1~\cite{Nosochkov:2025vmb} and were optimized to enhance interception of particles that would otherwise reach the restrictive apertures of the IR6 final-focus system. The selected locations provide relatively large betatron functions, approximately \SIrange{60}{90}{m}, and small dispersion in the relevant collimation plane, generally of order \SI{10}{cm} or less. The large betatron functions permit halo interception at comparatively large physical jaw half-apertures for a given setting in units of $\sigma_{\beta}$, thereby reducing sensitivity to mechanical positioning and alignment errors. The phase advances between the collimators and the limiting IR6 apertures were selected to enhance the interception of trajectories that would otherwise contribute to downstream losses.

\begin{figure*}[htbp]
\centering
\subfloat[\label{fig:ESR_Collimator_Drawing_1}]{\includegraphics[width=0.5\linewidth]{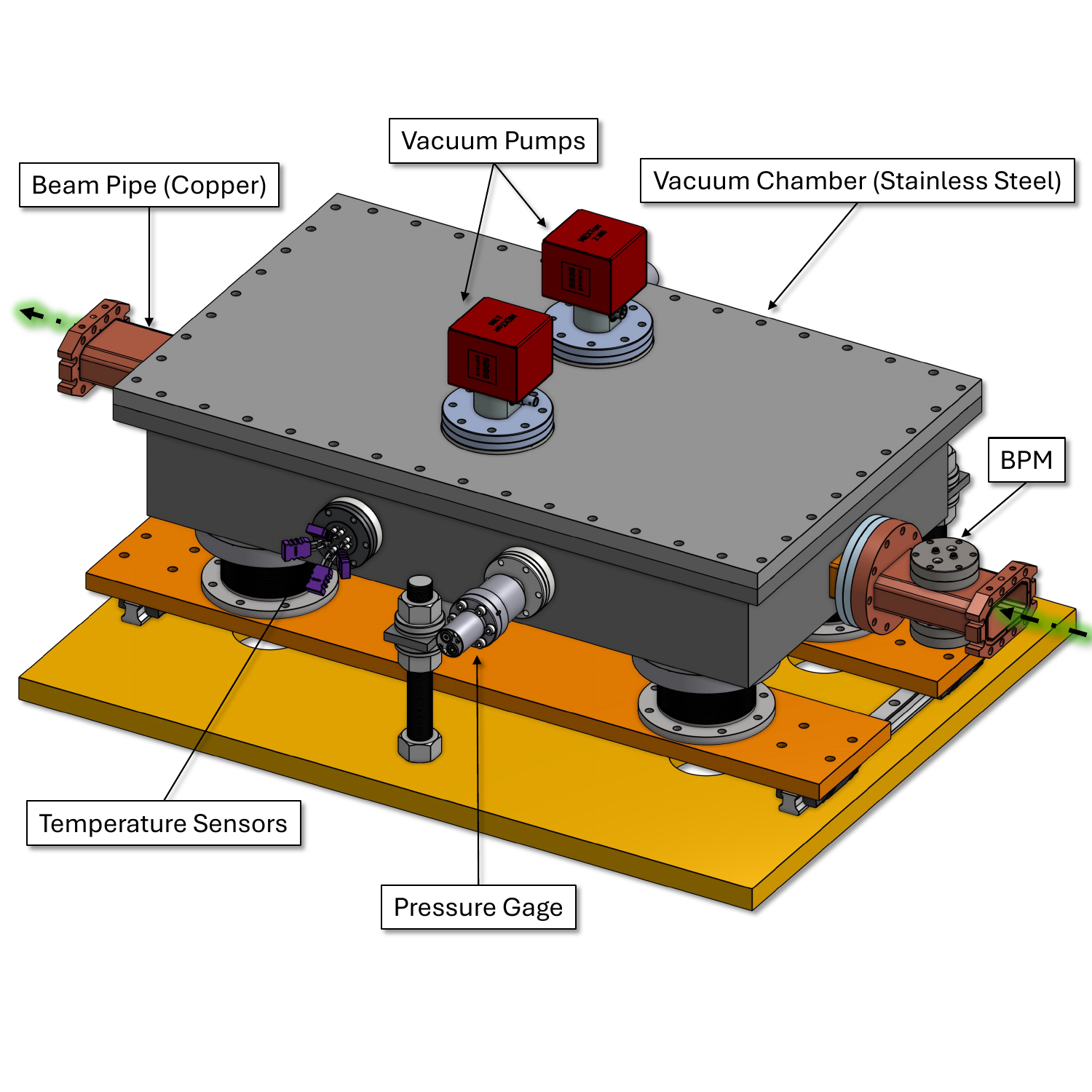}}
\hfill
\subfloat[\label{fig:ESR_Collimator_Drawing_2}]{\includegraphics[width=0.5\linewidth]{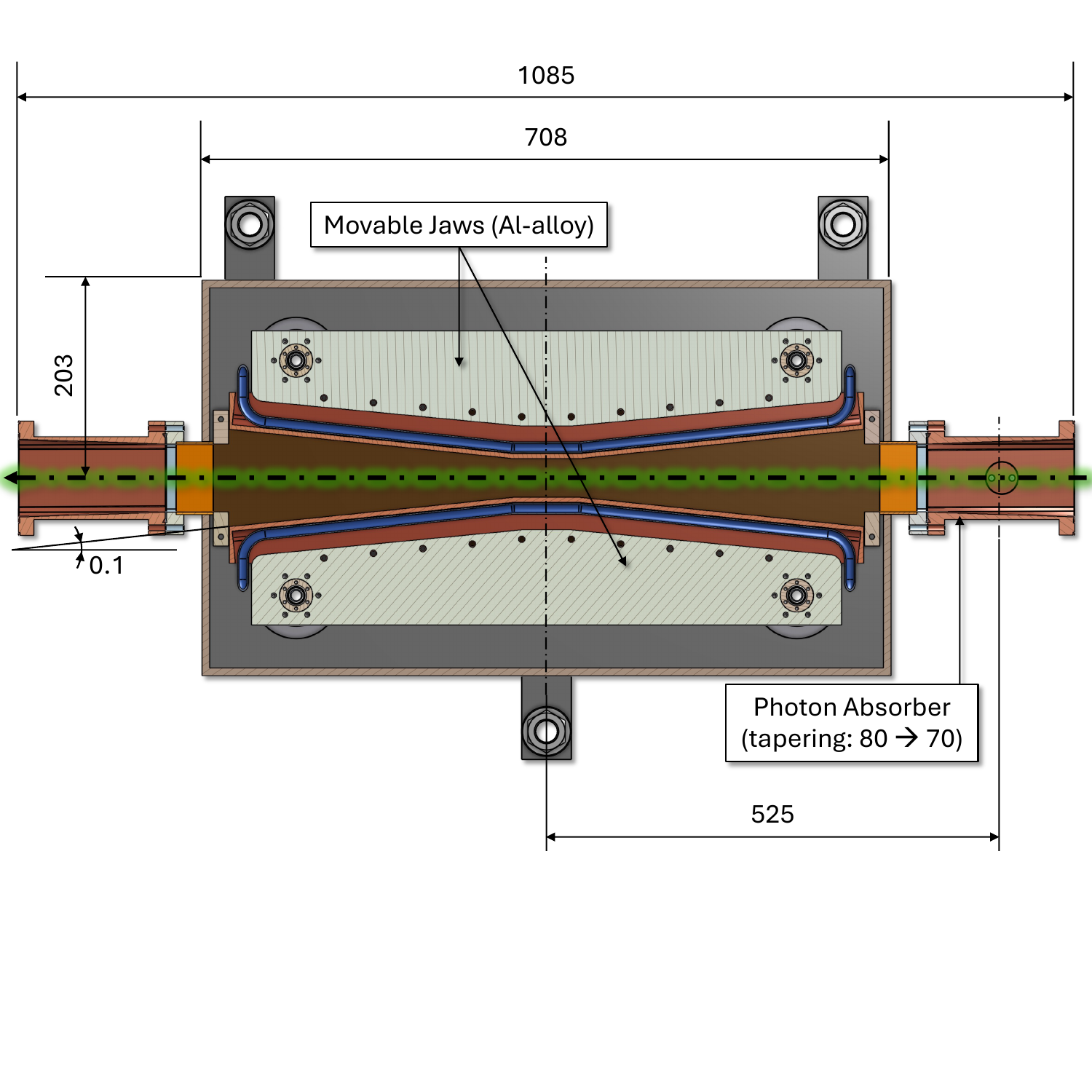}}
\\
\subfloat[\label{fig:ESR_Collimator_Drawing_3}]{\includegraphics[width=0.5\linewidth]{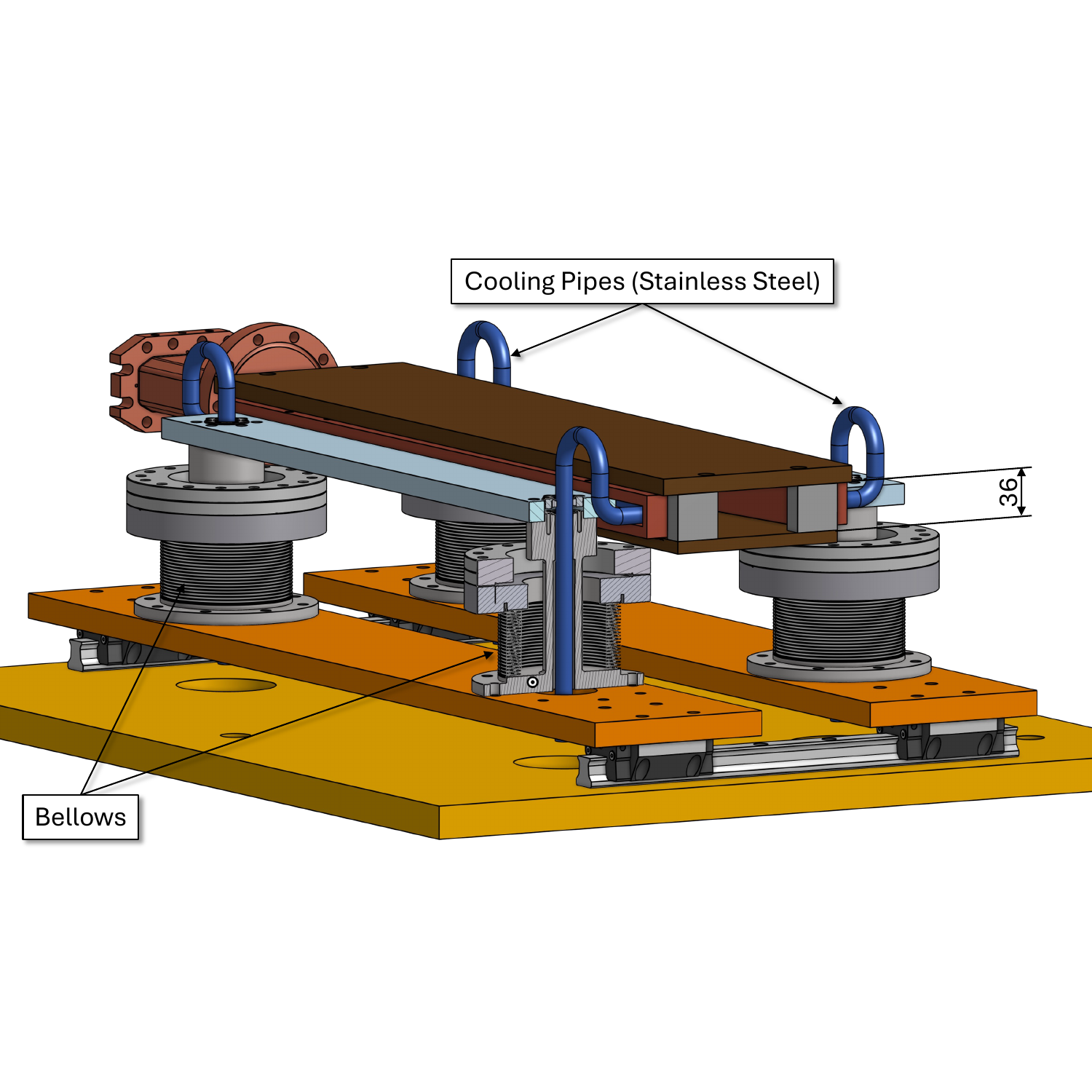}}
\hfill
\subfloat[\label{fig:ESR_Collimator_Drawing_4}]{\includegraphics[width=0.5\linewidth]{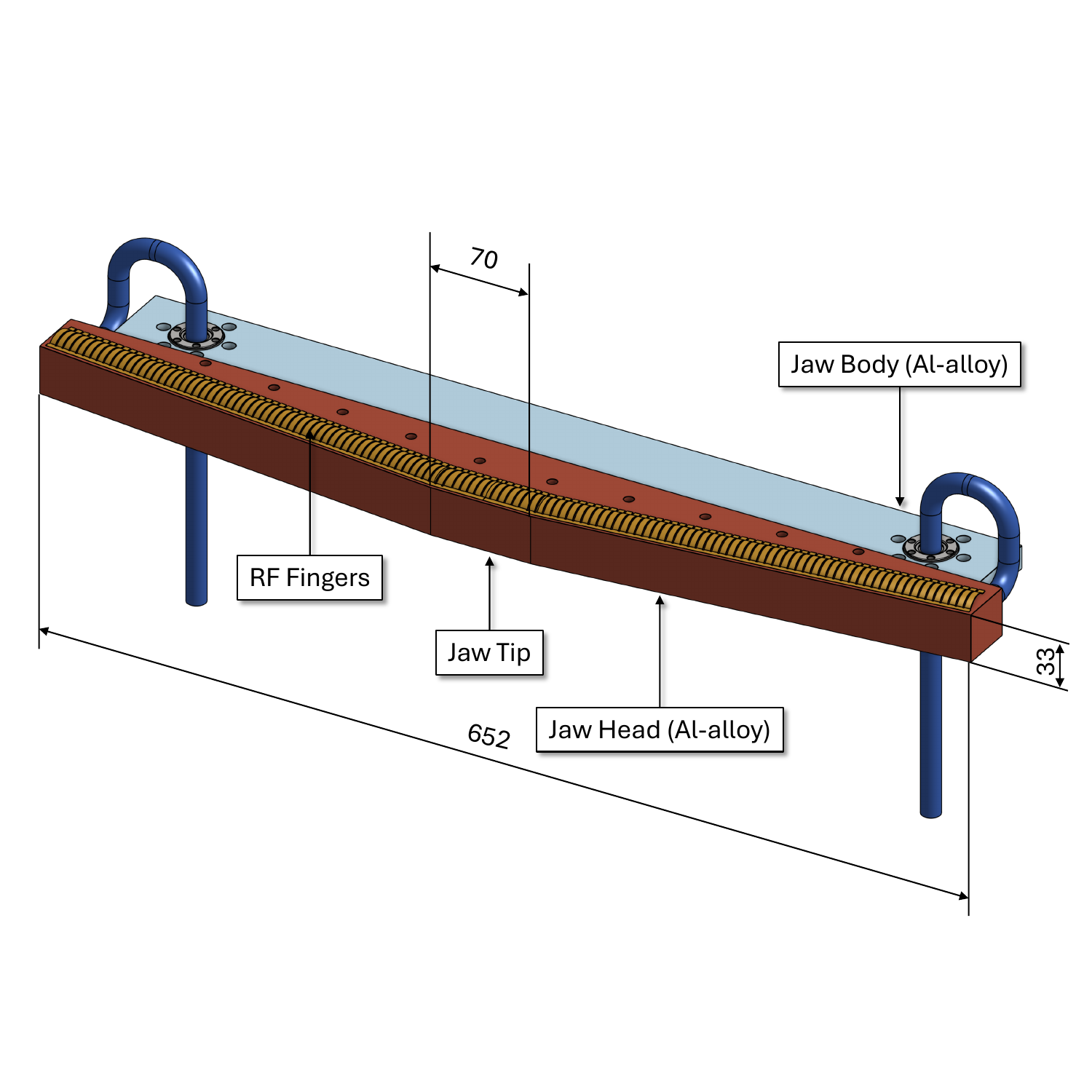}}
\caption{\label{fig:CollimatorDrawing}
Preliminary mechanical design of the ESR horizontal collimator. All linear dimensions are given in millimeters and angular dimensions in radians. (a) Collimator vacuum chamber. (b) Top view of the chamber and jaw assembly. (c) Movable jaw and support assembly. (d) Detailed geometry of the collimator jaw.}
\end{figure*}

For a given beam-energy configuration, the baseline optics study identifies one primary horizontal and one primary vertical collimator as the principal halo-intercepting devices. Additional collimator locations are retained to provide operational flexibility and to improve loss localization under more demanding beam-loss conditions. In particular, the optimized four-collimator configuration required for protection against fast instability-driven losses is discussed in Sec.~\ref{sec:machine_protection}.

Initial EIC operation is expected to focus primarily on 5 and \SI{10}{GeV} electron beams, whereas operation at \SI{18}{GeV} is foreseen for a later stage. The initial ESR collimation system is therefore planned with four installed collimator assemblies at the locations required for the 5 and \SI{10}{GeV} optics configurations. Additional hardware required for the dedicated \SI{18}{GeV} configuration can be installed as part of a future upgrade.

Figure~\ref{fig:CollimatorDrawing} shows the preliminary mechanical design of the ESR collimator assembly. Approximately \SI{2}{m} of longitudinal space is reserved at each collimator location, providing margin for installation and potential future modifications, while the present chamber design has a flange-to-flange length of approximately \SI{1}{m}. The mechanical concept draws on experience with movable collimators operated in high-current electron--positron colliders, including PEP-II~\cite{Seeman2000PEPIICollimator} and SuperKEKB~\cite{Ishibashi2017}. Each collimator employs opposing movable jaws whose tapered absorber heads can be positioned at the required beam half-aperture to intercept halo particles while providing a smooth transition between the nominal beam pipe and the restricted collimator aperture.

In the preceding system-level study~\cite{2v68-np87}, the collimators were represented by simplified tungsten absorbers in order to evaluate the achievable cleaning efficiency, downstream loss distributions, and beam-lifetime constraints. Optimization of the absorber material, effective jaw length, beam-impact survivability, thermomechanical response, and impedance was intentionally deferred. The optics and collimation configuration established in that study therefore serve as the reference system for the detailed design optimization developed in the following sections, beginning with the competing material, geometry, thermal, and impedance constraints discussed in Sec.~\ref{sec:design_constraints}.

\section{\label{sec:design_constraints}Design Constraints and Optimization Strategy}

The ESR collimator design is governed by several competing requirements involving beam-loss interception, beam-impact survivability, impedance, thermal performance, vacuum compatibility, and mechanical integration. The primary function of the system is to intercept beam-halo particles upstream of IR6 and thereby reduce losses at the restrictive apertures surrounding the ePIC detector and the SC final-focus magnets. Effective collimation requires sufficient absorber thickness to induce scattering and electromagnetic-shower development so that intercepted particles are either absorbed locally or removed from the machine acceptance before reaching sensitive downstream components.

The absorber material must simultaneously provide adequate robustness against both routine halo losses and transient abnormal beam impacts. For a given incident beam distribution, the material radiation length, density, and thermophysical properties determine the development of the electromagnetic shower and the spatial distribution of deposited energy. Materials with short radiation lengths permit compact absorbers and efficient shower development, but can also concentrate the deposited energy within a relatively small volume. Materials with longer radiation lengths generally distribute the energy over a larger volume and can therefore reduce the peak energy-deposition density, although at the cost of a longer absorber. The resulting temperature rise and thermomechanical response additionally depend on the specific heat capacity, thermal conductivity, coefficient of thermal expansion, elastic properties, and mechanical strength of the material.

The collimator geometry and electrical properties also contribute directly to the ESR beam-coupling impedance. The small operational half-apertures required for efficient halo interception enhance both geometric and resistive-wall wakefields, while the finite electrical conductivity of the jaw material contributes to beam-induced heating. The absorber geometry must therefore provide sufficient interaction length for effective collimation without introducing unnecessary material close to the beam. Additional engineering constraints include manufacturability, dimensional stability, cooling, ultrahigh-vacuum (UHV) compatibility, operational reliability, and integration into a movable-jaw assembly.

For an otherwise fixed geometry in the resistive-wall regime, the impedance contribution increases approximately with the longitudinal extent of the conducting surface and decreases with increasing electrical conductivity~\cite{Chao2023Handbook}. The absorber material and effective tip length are therefore coupled design parameters. A material with relatively low electrical conductivity can impose a significant resistive-wall impedance and beam-induced heating penalty, whereas extending the absorber beyond the length required to achieve the desired downstream loss suppression increases the conducting surface exposed to the beam without necessarily providing a corresponding improvement in collimation performance. Determining the minimum effective absorber length is therefore an important part of the optimization.

Vacuum compatibility imposes an additional constraint on the choice of engineering alloys. Materials used in the ESR collimator must be compatible with the UHV environment and with the thermal and vacuum-processing procedures foreseen for the storage ring. Particular attention was given to alloying elements with comparatively high vapor pressure, which may increase the risk of outgassing or material redistribution during elevated-temperature operation or abnormal beam-heating events. For the present material screening, aluminum alloys containing more than approximately \SI{1}{\percent} Zn by weight were therefore excluded from consideration. This criterion is used here as an engineering constraint for the ESR collimator design rather than as a general material limit for UHV accelerator components.

The optimization is consequently performed as a coupled problem rather than by maximizing any single material or geometric property. Monte Carlo particle-transport simulations are first used to determine the absorber length required for adequate suppression of downstream losses and to calculate the three-dimensional energy-deposition distributions produced by beam impacts. These distributions are then used to compare the transient thermal response and survivability of candidate absorber materials. Electromagnetic wakefield calculations quantify the corresponding impedance contribution and beam-induced power deposition, while finite-element thermomechanical calculations assess the steady-state response of the selected jaw design. The combined results are used to identify a practical compromise among collimation performance, impact tolerance, impedance, thermal performance, vacuum compatibility, and mechanical integration.

\section{\label{sec:length_optimization}Collimator Length Optimization}

The system-level ESR collimation studies reported in Ref.~\cite{2v68-np87} showed that primary collimators located in IR4 can reduce charged-particle losses in the interaction region by one to two orders of magnitude for the beam-loss sources considered. Those studies employed simplified tungsten absorbers and did not include a systematic determination of the minimum effective jaw length, $L$. Because the absorber length affects the available mechanical envelope, the volume of material exposed to beam impacts, and both the geometric and resistive-wall impedance, identifying the shortest length that provides adequate downstream loss suppression is an important design objective.

A dedicated length scan was performed using the ESR multi-turn particle-tracking framework described in Ref.~\cite{2v68-np87}. The results presented in this section correspond to the \SI{10}{GeV} electron-beam configuration. Initial particle distributions generated by Touschek scattering and beam--gas interactions were tracked with Xsuite~\cite{Iadarola:2023fuk}. Particle interactions in the collimator material, including electromagnetic-shower production, were modeled through the Xsuite--BDSIM interface~\cite{NEVAY2020107200} using Geant4~\cite{AGOSTINELLI2003250,1610988,ALLISON2016186}. Surviving primary particles and secondary particles emerging from the collimator were subsequently tracked through the downstream ESR lattice.

To reduce the computational cost of the multi-turn calculation, downstream transport of secondary particles was restricted to charged particles with energies above \SI{1}{GeV}. Lower-energy charged shower products were found to be lost predominantly in the vicinity of the collimator or in the arc between IR4 and IR6. Photons and other neutral secondary particles were not propagated through the downstream lattice in this scan. The quantity evaluated below therefore represents the suppression of transported charged-particle losses in IR6 and should not be interpreted as a complete calculation of electromagnetic-shower containment, neutral-particle transport, or detector backgrounds.

The length scan considered representative high-$Z$, intermediate-$Z$, and low-$Z$ absorber materials: tungsten, tantalum, copper, titanium, aluminum, molybdenum--graphite composite (MoGr), copper--diamond composite (CuCD), and carbon-fiber composite (CFC). Their radiation lengths and other material properties are summarized in Table~\ref{tab:MaterialProperties}. For each material, the absorber length was varied while the resulting charged-particle loss rate was integrated over the IR6 final-focus cryostat region, extending from $z=-\SI{15}{m}$ to $z=+\SI{22.5}{m}$ relative to IP6.

The radiation length, $X_0$, provides a natural scale for comparing the initial development of electromagnetic cascades in materials with different atomic numbers and densities. It is conventionally defined as the mean distance over which a high-energy electron loses all but a fraction $e^{-1}$ of its energy through bremsstrahlung ~\cite{PDG2024}. The scan results are therefore expressed as a function of the normalized absorber length, $L/X_0$. High-$Z$, high-density materials generally have short radiation lengths and consequently require smaller physical lengths to produce a given amount of bremsstrahlung and scattering, whereas low-$Z$ materials require longer absorbers.

The use of $L/X_0$ as the normalization variable does not imply that an absorber with $L\simeq X_0$ fully contains the electromagnetic shower. Complete calorimetric containment of a \SI{10}{GeV} electromagnetic shower would require substantially more material. In the present optimization, the relevant question is instead whether the absorber produces sufficient interaction and scattering to remove the intercepted electrons and energetic charged secondaries from the downstream machine acceptance.

\begin{figure}[htbp]
\centering
\includegraphics[width=\linewidth]{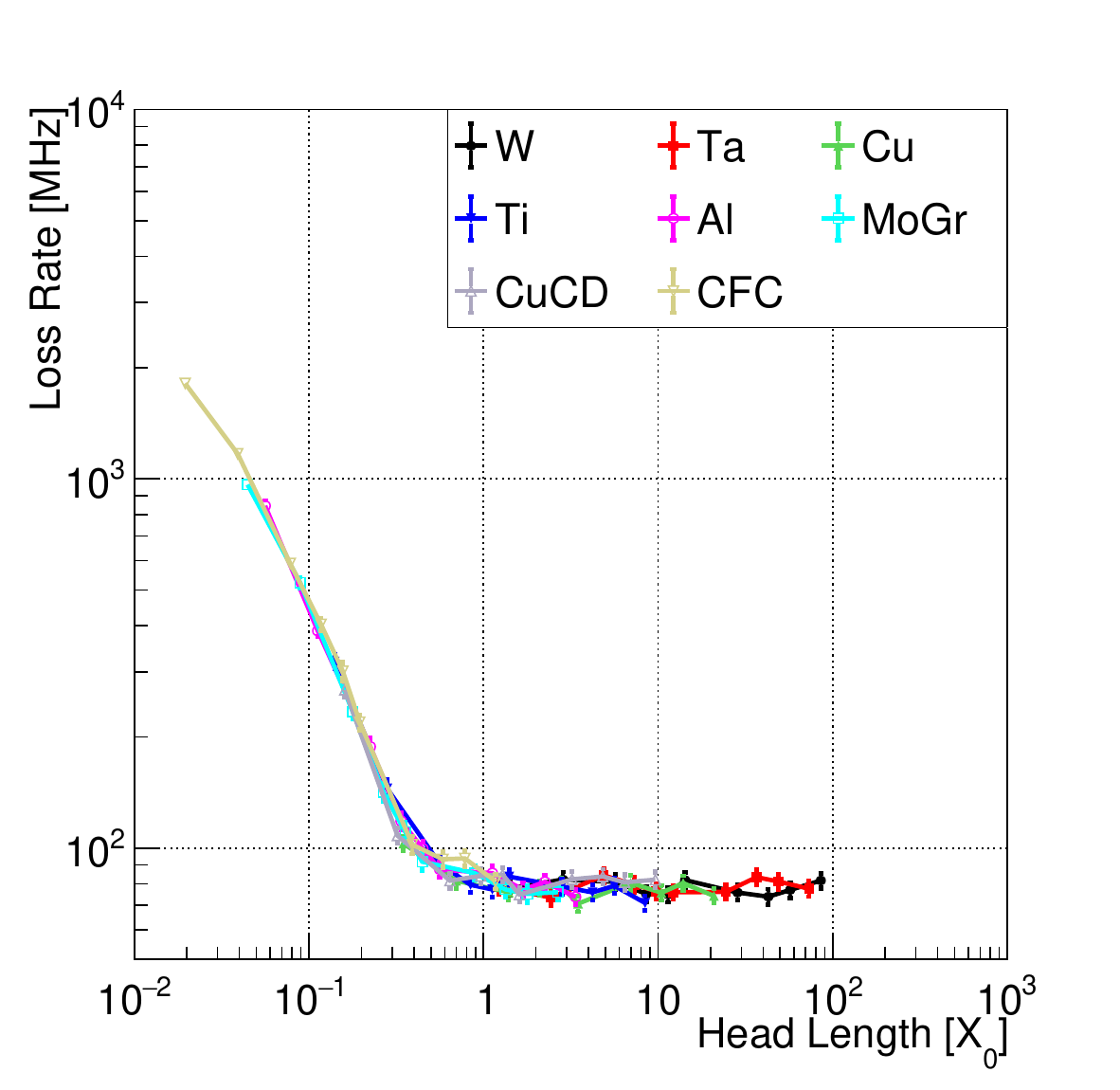}
\caption{\label{fig:IRlossesVsTipLength_10gev}
Integrated loss rates in the IR6 final-focus cryostat region, from $z=-\SI{15}{m}$ to $z=+\SI{22.5}{m}$ relative to IP6, as a function of the collimator absorber length normalized by the material radiation length at \SI{10}{GeV}.}
\end{figure}

Figure~\ref{fig:IRlossesVsTipLength_10gev} shows that the normalized length dependence is approximately common to the materials considered, despite their substantially different physical radiation lengths. The IR6 charged-particle loss rate decreases rapidly as the absorber length is increased from a small fraction of $X_0$ to approximately one radiation length. Beyond $L/X_0\simeq 1$, the additional reduction is comparatively small and the loss rates approach broad plateaus.

These results indicate that an absorber length of roughly one radiation length is sufficient to approach the minimum transported charged-particle losses for the ESR interaction region within the study's tracking model and metric, which excludes photons and other neutral secondaries. A length of approximately $X_0$ was therefore adopted as an initial screening value for the subsequent comparison of candidate absorber materials, rather than as a rigid requirement on the final jaw geometry.

The final absorber length is refined in Sec.~\ref{sec:head_refinement} using the selected Al2219-T62 material, the updated four-collimator configuration, and design-specific beam-impact calculations. That analysis shows that the simulated IR6 charged-particle loss rate has already approached its plateau for Al2219-T62 tip lengths above approximately \SI{60}{mm}, leading to the selection of a \SI{70}{mm} tapered tip, corresponding to approximately $0.8\,X_0$, as the baseline engineering design.

\section{\label{sec:survivability}Temperature Rise and Survivability Studies}

\begin{table*}[tb]
\caption{\label{tab:MaterialProperties}
Candidate collimator jaw materials considered in this study together with selected physical, thermal, and mechanical properties. Radiation lengths $X_0$ and densities $\rho$ were obtained from the Particle Data Group (PDG) material database~\cite{PDGAtomicNuclearProperties}. Mechanical and thermophysical properties of elemental materials were compiled from the MatWeb materials database~\cite{MatWeb}. Properties of engineering alloys, composites, and advanced collimator materials were compiled from manufacturer datasheets and the references listed in Table~\ref{tab:MaterialSources}. Here, CFC denotes carbon-fiber composite, MoGr denotes a molybdenum-graphite composite, and CuCD denotes a copper-diamond composite reinforced with chromium and zirconium.}
\centering
\scriptsize
\resizebox{\textwidth}{!}{
\begin{tabular}{lcccccccccccccc}
\hline\hline
Material &
$\rho$ &
$X_0$ &
$E$ &
$\alpha\times10^{-6}$ &
$\nu$ &
$\sigma$ &
$\Delta T_{\rm shock}^{\rm elastic}$ &
Plasticity &
$\Delta T_{\rm shock}^{\rm practical}$ &
Failure &
$T_m/T_s$ &
$\Delta H_f$ &
$L_f=\rho\Delta H_f$ &
$k$
\\
&
[g/cm$^3$] &
[mm] &
[GPa] &
[K$^{-1}$] &
&
[MPa] &
[$^\circ$C] &
Factor &
[$^\circ$C] &
Mode &
[K] &
[J/g] &
[J/cm$^3$] &
[W/m/K]
\\
\hline

W
& 19.30 & 3.50 & 400 & 4.4 & 0.28 & 980
& 401 & $\times1$ & 401
& Cracking
& 3643 & 184.2 & 3546 & 163.3\\

Ta
& 16.65 & 4.09 & 186 & 6.5 & 0.35 & 485
& 261 & $\times2$ & 522
& Plastic flow
& 3269 & 170.0 & 2837 & 54.4\\

Cu
& 8.96 & 14.36 & 110 & 16.4 & 0.343 & 210
& 76 & $\times2$ & 152
& Plastic flow
& 1357 & 204.8 & 1835 & 385.0\\

CuCD
& 5.40 & 31.10 & 160 & 6.56 & 0.07 & 104
& 92 & $\times1$ & 92
& Interface cracking
& 1357 & 204.8 & 1106 & 337.9\\

Ti
& 4.51 & 35.60 & 116 & 8.9 & 0.34 & 220
& 141 & $\times2$ & 282
& Plastic flow
& 1933 & 435.4 & 1964 & 17.0\\

Ti Grade 5
& 4.43 & 36.2 & 113.8 & 9.7 & 0.342 & 950
& 566 & $\times2$ & 1132
& Plastic flow
& 1877 & 435.4 & 1929 & 6.7\\

Ti Grade 9
& 4.48 & 35.8 & 100.0 & 9.97 & 0.30 & 620
& 435 & $\times2$ & 870
& Plastic flow
& 1973 & 435.4 & 1951 & 8.3\\

Al
& 2.70 & 88.97 & 68.0 & 24.0 & 0.33 & 90
& 37 & $\times2$ & 74
& Plastic flow
& 934 & 386.9 & 1045 & 201.0\\

Al6061-T6
& 2.70 & 89.0 & 68.9 & 23.6 & 0.33 & 310
& 128 & $\times2$ & 256
& Plastic flow
& 855 & 386.9 & 1045 & 167.0\\

Al2219-T62
& 2.84 & 84.6 & 73.1 & 22.3 & 0.33 & 414
& 170 & $\times2$ & 340
& Plastic flow
& 816 & 386.9 & 1099 & 120.0\\

MoGr
& 2.57 & 112.20 & 60 & 2.7 & 0.10 & 60
& 333 & $\times1$ & 333
& Matrix cracking
& 3900 & $\sim60000$ & 154200 & 706.2\\

CFC
& 1.67 & 255.70 & 56 & 3.9 & 0.20 & 165
& 604 & $\times1$ & 604
& Matrix cracking
& 3900 & $\sim60000$ & 100200 & 197.0\\

\hline\hline
\end{tabular}
}
\end{table*}

\begin{table*}[htbp]
\caption{\label{tab:MaterialSources}
Sources used for material properties.
}
\centering
\begin{tabular}{r|l|l}
\hline\hline
Material & Composition & Source \\
\hline
W, Ta, Cu, Ti, Al & Pure (100\%) & MatWeb~\cite{MatWeb},  \\
Al6061-T6 & Al/Mg/Si/Cr/Cu/Fe/Mn/Ti/Zn & ASM-MatWeb~\cite{AsmMatWeb} \\
& 97.36/01.01/00.61/0.20/0.30/00.30/0.051/0.051/0.101\% & \\
Al2219-T62 & Al/Cu/Mn/Fe/Si/Ti/V/Zr/Zn/Mg &  \\
& 92.36/6.39/0.30/0.20/0.20/0.10/0.10/0.20/0.10/0.0499\% & \\
Ti Grade 5 & Ti/Al/V/Fe/O/N/C/H &  \\
& 89.54/5.97/3.98/0.199/0.199/0.0497/0.0497/0.00995\% & \\
Ti Grade 9 & Ti/Al/V/Fe/O/C/N/H &  \\
& 93.86/3.06/2.55/0.204/0.204/0.0816/0.0306/0.0102\% & \\
\hline
MoGr & Mo/C/Ti & CERN FCC studies~\cite{GUARDIAVALENZUELA201872,Redaelli2019TCSPM} \\
& 2.01/97.80/0.19\% & \\
CuCD & Cu/C/Cr/Zr &  \\
& 26.96/72.69/0.33/0.02\% & \\
\hline
CFC & C & CFCCARBON material datasheet~\cite{CFCCarbonCFC}, \\
& Pure (100\%) & Radiation damage studies~\cite{SIMOS2021100028}, \\
& & CERN LHC studies~\cite{Antonakakis2012TCDQ} \\
\hline\hline
\end{tabular}
\end{table*}

The response of a collimator to a rapid beam impact is governed by the spatial and temporal distributions of deposited energy and by the thermophysical and mechanical properties of the absorber. For the fast beam-loss scenarios considered here, the peak energy-deposition density and resulting local temperature rise provide useful first-order measures of impact severity. Because material failure may also depend on stress-wave propagation, strain-rate effects, plastic deformation, and fracture, the calculated temperature distributions are interpreted using engineering screening criteria rather than as exact failure thresholds.

At high deposited-energy densities, a constant-specific-heat approximation becomes inaccurate because the specific heat varies with temperature and phase transitions may occur. The calculations therefore employ an enthalpy-based energy balance that includes temperature-dependent specific heat and the latent heat associated with melting or sublimation. The resulting temperature distributions are used to estimate the material volumes exceeding the adopted thermal-shock criterion and the relevant phase-transition threshold. The material properties and data sources are summarized in Tables~\ref{tab:MaterialProperties} and~\ref{tab:MaterialSources}.

The analysis framework is assessed through two complementary benchmarks. The Geant4 energy-deposition calculations and scoring-resolution requirements are first compared with published International Linear Collider (ILC) spoiler simulations. The temperature-rise and thermal-diffusion treatment is then compared with published Advanced Photon Source Upgrade (APS-U) simulations and beam-abort experiments. These comparisons provide benchmark support for the electromagnetic-shower and temperature-rise calculations, but do not constitute a complete validation of the transient elastoplastic or hydrodynamic material response.

The benchmarked framework is subsequently applied to candidate ESR absorber materials under defined reference impact conditions. The objective is to compare their thermal response, estimate model-dependent thresholds for thermal-shock-related damage and phase transitions, and evaluate the tradeoffs among impact tolerance, absorber length, thermal and electrical properties, vacuum compatibility, and mechanical integration. The resulting estimates guide the selection of the baseline absorber material and inform the associated machine-protection requirements.

\subsection{\label{sec:TemperatureRiseCalculation}Enthalpy-Based Temperature Rise Calculation}

Material properties used throughout this study were compiled from the PDG material database, engineering handbooks, manufacturer datasheets, and published collimator R\&D studies. A summary of the adopted parameters and their sources is provided in Tables~\ref{tab:MaterialProperties} and~\ref{tab:MaterialSources}. In addition to density and radiation length, the data include thermal, elastic, and mechanical properties relevant to beam-impact survivability.

A commonly used figure of merit for material robustness is the thermal-shock limit, defined as the temperature increase for which the thermally induced stress reaches a characteristic material-strength limit. Assuming a constrained isotropic material, the elastic thermal-shock limit can be estimated as~\cite{Kingery1955}
\begin{equation}
\Delta T_{\mathrm{shock}}^{\mathrm{elastic}}
=
\frac{\sigma(1-\nu)}{E\alpha},
\label{eq:thermalShock}
\end{equation}
where $\sigma$ is the selected material-strength parameter, $E$ is Young's modulus, $\nu$ is Poisson's ratio, and $\alpha$ is the coefficient of thermal expansion. In the present study, $\sigma$ is taken as the ultimate tensile strength (UTS) listed in Table~\ref{tab:MaterialProperties}. The resulting quantity is not interpreted as the temperature rise for the onset of yielding or the first occurrence of permanent plastic deformation. Instead, it is used as an engineering indicator of the thermal-shock tolerance associated with severe mechanical loading, consistent with the survivability objective of the present study.

The distinction is important for collimator applications. The onset of yielding does not necessarily imply loss of functionality during an accidental beam impact, since localized plastic deformation may be tolerable provided that the absorber remains structurally intact and capable of performing its beam-protection function. Conversely, the UTS-based estimate does not represent a universal fracture or failure criterion. The actual response depends on the spatial and temporal stress distribution, plastic stress redistribution, strain-rate effects, stress-wave propagation, geometry, and boundary conditions. The UTS-based criterion adopted here should therefore be regarded as a screening-level measure of the transition from elastic thermal response toward severe mechanical loading, rather than as a prediction of a uniquely defined material-failure point.

For ductile metallic materials, thermal-shock tolerance depends not only on the elastic thermal stress but also on the material's ability to accommodate thermally induced strains through plastic deformation. Plastic deformation can redistribute transient thermal stresses and allow additional strain energy to be absorbed before cracking or other severe structural damage occurs~\cite{Manson1954,Au1967,Bertarelli2016,MA2023101353}. Consequently, a purely elastic thermal-shock estimate does not necessarily represent the practical beam-impact tolerance of a ductile collimator material.

To represent this additional deformation capacity within the simplified screening framework used in the present study, a \textit{practical thermal-shock threshold} is defined as twice the UTS-based elastic estimate for the ductile metallic materials considered here, namely Ta, Ti, Cu, and Al alloys. This factor is introduced as an empirical engineering correction and should not be interpreted as a universal material-failure criterion or as a substitute for a transient elastoplastic analysis. Its applicability to beam-impact conditions is assessed against the Advanced Photon Source Upgrade (APS-U) beam-impact experiments discussed in Sec.~\ref{sec:APS_U_benchmark}. For W and the brittle/composite materials considered here, namely MoGr, CuCD, and CFC, the elastic estimate is retained without this correction, consistent with their cracking-dominated failure mode (Table~\ref{tab:MaterialProperties}). Commercial tungsten typically exhibits a ductile-to-brittle transition temperature in the range \SIrange{250}{450}{\celsius}~\cite{Bertarelli2016}, so a room-temperature-baseline screening estimate should not assume the plastic stress-redistribution capacity underlying the ductile correction.

In addition to the thermal-shock criterion, phase transitions provide an independent limit on the material response. For metallic absorbers, the relevant threshold is the melting temperature $T_m$, while for graphite-based materials the corresponding threshold is the sublimation temperature $T_s$. Exceeding these temperatures indicates irreversible material transformation and therefore provides a complementary measure of survivability. Both the thermal-shock criterion and the phase-transition threshold are used in the analysis below.

For a deposited energy density $E_{\mathrm{dep}}$, the temperature rise under a constant-specific-heat approximation is
\begin{equation}
\Delta T = \frac{E_{\mathrm{dep}}}{\rho c_p},
\label{eq:dTsimple}
\end{equation}
where $E_{\mathrm{dep}}$ is expressed in J/cm$^3$, $\rho$ is the material density in g/cm$^3$, and $c_p$ is the specific heat capacity in J/(g\,K). Equation~\eqref{eq:dTsimple} is adequate for preliminary estimates when $c_p$ is approximately constant, but becomes increasingly inaccurate as the temperature approaches the range where the material properties vary substantially or a phase transition occurs.

Thermal transport during the beam-loss event provides an additional modification to the instantaneous energy-deposition profile. For very rapid losses, the deposited energy remains approximately localized during the energy-deposition process. For losses extending over multiple machine turns, however, thermal diffusion can redistribute part of the deposited energy and reduce the peak energy density. This effect is particularly relevant when comparing calculations with experimental beam-impact data obtained under multi-turn beam-abort conditions.

The importance of this effect has been demonstrated by beam-impact studies at the APS-U, where the observed damage threshold depends not only on the total deposited energy but also on the temporal evolution of the beam loss~\cite{BorlandNAPAC2019,Dooling2022}. The present work therefore incorporates a first-order thermal-diffusion correction based on the approach used in the APS-U beam-abort studies~\cite{BorlandNAPAC2019}.

The characteristic thermal diffusion length is estimated as
\begin{equation}
\sigma_{\rm diff}=\sqrt{D_{\rm diff}t_{\rm loss}},
\end{equation}
where
\begin{equation}
D_{\rm diff}=\frac{k}{\rho c_p(T_0)}
\end{equation}
is the thermal diffusivity, $k$ is the thermal conductivity, $\rho$ is the material density, and $c_p(T_0)$ is the specific heat capacity evaluated at the initial temperature $T_0=\SI{300}{K}$. Room-temperature material properties are used in this estimate, providing a first-order approximation of heat transport during the beam-loss event. Although the thermal diffusivity varies with temperature, this approximation was found to be adequate for the benchmark studies presented below and avoids the computational cost of a fully coupled transient heat-conduction calculation.

Assuming approximately Gaussian transverse beam profiles, thermal diffusion is represented as an additional broadening of the effective transverse energy-deposition distribution. The effective transverse dimensions are therefore approximated as
\begin{equation}
\sigma_{x,\rm eff}
=
\sqrt{\sigma_x^2+\sigma_{\rm diff}^2},
\qquad
\sigma_{y,\rm eff}
=
\sqrt{\sigma_y^2+\sigma_{\rm diff}^2},
\end{equation}
where $\sigma_x$ and $\sigma_y$ are the RMS transverse beam sizes at the impact location. Conservation of the deposited energy then gives the corresponding correction to the peak energy density,
\begin{equation}
E_{\rm dep}^{\rm corr}
=
E_{\rm dep}
\frac{\sigma_x\sigma_y}
{\sqrt{\sigma_x^2+\sigma_{\rm diff}^2}
 \sqrt{\sigma_y^2+\sigma_{\rm diff}^2}}.
\label{eq:diffusionScaling}
\end{equation}

The importance of this correction depends strongly on the relative magnitude of the thermal diffusion length and the beam dimensions. When $\sigma_{\rm diff} \ll \sigma_x,\sigma_y$, the correction becomes negligible and the deposited energy remains effectively confined within the original beam footprint. Conversely, when $\sigma_{\rm diff}$ becomes comparable to the beam dimensions, thermal diffusion substantially broadens the effective deposited-energy distribution and reduces the peak energy density. Since beam-induced damage is governed primarily by the local peak energy density rather than the total deposited energy, even a modest increase in the effective beam size can significantly alter the predicted onset of mechanical damage or melting.

This behavior is particularly important for highly focused electron beams, where the thermal diffusion length during a microsecond-scale beam loss may become comparable to the transverse beam dimensions. In such cases, neglecting thermal diffusion may substantially overestimate the peak temperature rise and consequently underestimate the survivability margin. Conversely, for larger beam sizes, the correction is generally more modest. Nevertheless, it remains important when comparing simulation results with experimentally observed damage thresholds and when evaluating survivability margins near material failure limits.

The final temperature is obtained using an enthalpy-based integration that accounts for the temperature dependence of the specific heat capacity and the energy required for phase transitions:
\begin{equation}
E_{\mathrm{dep}}^{\mathrm{corr}}
=
\int_{T_0}^{T_f}\rho c_p(T)\,dT + L_f,
\label{eq:enthalpy}
\end{equation}
where $T_f$ is the final temperature and $L_f$ represents the latent energy associated with the relevant phase transition. For metallic materials, $L_f$ is the volumetric latent heat of fusion, while for graphite-based materials the corresponding sublimation energy is used. The final temperature is obtained numerically by integrating the deposited energy using the temperature-dependent specific heat data.

\begin{figure}[htbp] 
\centering 
\includegraphics[width=\linewidth]{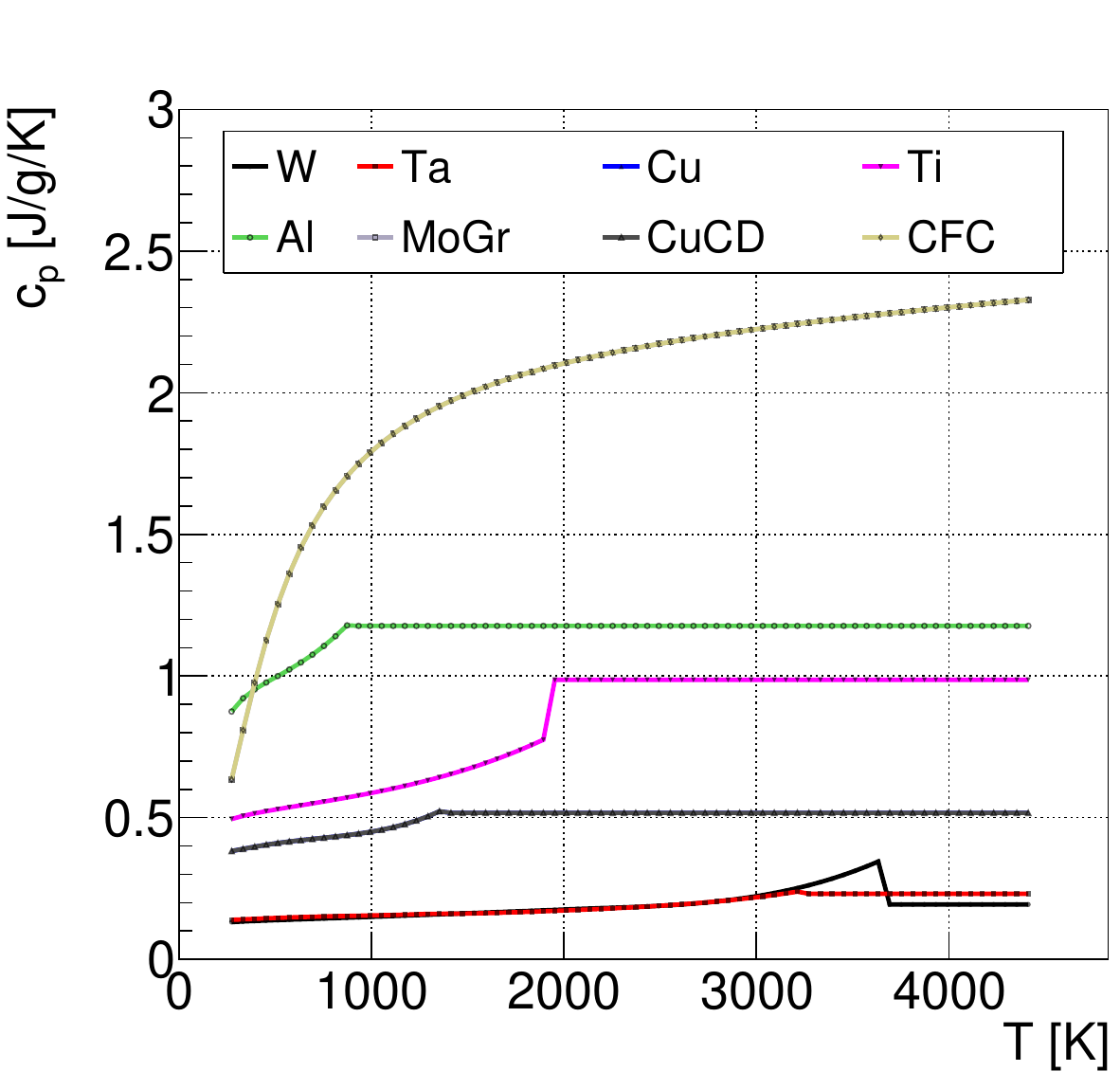} 
\caption{\label{fig:HeatCapacity} Temperature-dependent specific heat capacity $c_p$ used in the enthalpy-based temperature rise calculations for the candidate collimator materials. Thermodynamic data were obtained from the NIST Chemistry WebBook~\cite{NISTWebBook}. The specific heat capacities of Al, Ti, and Cu alloys were approximated using the corresponding pure-material data, while graphite data were used for both CFC and MoGr. } 
\end{figure} 

The temperature-dependent specific heat capacities are obtained from thermodynamic data in the NIST Chemistry WebBook~\cite{NISTWebBook}. Figure~\ref{fig:HeatCapacity} shows the data used in the calculations. For Al, Ti, and Cu alloys, the corresponding pure-material data are used as an approximation, while graphite data are used for both CFC and MoGr.

For metallic materials, the volumetric energy density associated with complete melting is calculated as
\begin{equation}
L_f=\rho\Delta H_f,
\end{equation}
where $\Delta H_f$ is the specific enthalpy of fusion. The corresponding values are listed in Table~\ref{tab:MaterialProperties}. This enthalpy-based treatment provides a more realistic estimate of the temperature response at large deposited-energy densities than a constant-specific-heat approximation.

The combined Monte Carlo energy-deposition, thermal-diffusion, and enthalpy-based temperature-rise methodology is benchmarked in the following sections against published ILC simulations and APS-U beam-impact results. The ILC comparison tests the energy-deposition calculation, while the APS-U comparison provides an experimental test of the thermal response and the thermal-diffusion correction under multi-turn beam-loss conditions.

\subsection{\label{sec:IlcBenchmarking}Validation Against the ILC Spoiler Study}

Before applying the temperature-rise methodology to the ESR collimator system, the simulation framework was benchmarked against published International Linear Collider (ILC) spoiler survivability studies~\cite{FernandezHernando2006}. These benchmark cases were selected because detailed FLUKA and Geant4 reference calculations are available in the literature, providing a well-established basis for validating both the energy-deposition and temperature-rise calculations.

\begin{figure*}[htbp]
\centering
\includegraphics[width=\linewidth]{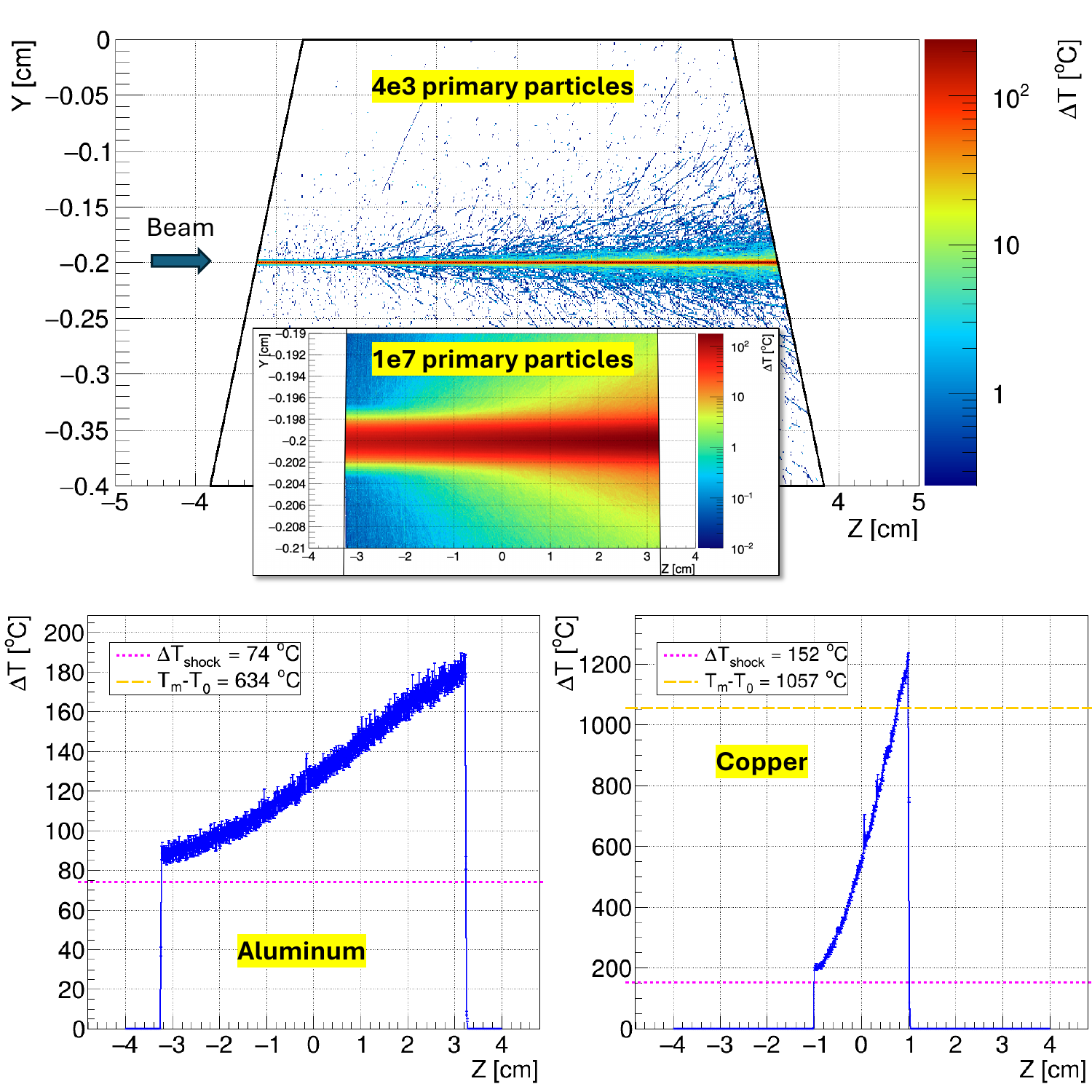}
\caption{\label{fig:ILC_Benchmark}
Geant4 benchmark simulations of ILC spoiler survivability following the configuration of Ref.~\cite{FernandezHernando2006}. A bunch of $2\times10^{10}$ electrons with an energy of \SI{250}{GeV} impacts tapered aluminum and copper spoiler jaws at a vertical offset of \SI{2}{mm} from the jaw edge. Energy deposition was scored using voxels of size $10\times1\times100~\si{\micro\meter^3}$. Top: two-dimensional temperature-rise maps in the aluminum spoiler for different Monte Carlo statistics. Bottom left: temperature rise along the aluminum spoiler jaw. Bottom right: temperature rise along the copper spoiler jaw. The horizontal magenta dotted line indicates the thermal-shock limit, while the horizontal orange dashed line denotes the melting temperature-rise threshold. The calculated peak temperature rises are in good agreement with the published ILC results of Ref.~\cite{FernandezHernando2006}, which concluded that a single ILC bunch exceeds the melting threshold of copper and the fracture threshold of aluminum.
}
\end{figure*}

The benchmark configuration was implemented to reproduce the ILC spoiler study of Ref.~\cite{FernandezHernando2006}. The simulated geometry consisted of a tapered aluminum or copper spoiler with a taper angle of \SI{335}{mrad} and a spoiler-tip length of $0.6X_0$. A bunch of $2\times10^{10}$ electrons with an energy of \SI{250}{GeV} was incident on the spoiler with transverse beam sizes of $\sigma_x=\SI{111}{\micro\meter}$ and $\sigma_y=\SI{9}{\micro\meter}$, and a vertical beam offset of \SI{2}{mm} relative to the spoiler tip. A total of $10^7$ primary electrons were simulated to ensure adequate statistical precision.

The benchmark simulations were performed using the same Geant4-based energy-deposition framework employed throughout this work. To enable a direct comparison with the published results, the temperature-rise calculation followed the methodology of Ref.~\cite{FernandezHernando2006}, employing a constant specific-heat approximation with $c_p=\SI{0.899}{J\,g^{-1}\,K^{-1}}$ for aluminum and $c_p=\SI{0.385}{J\,g^{-1}\,K^{-1}}$ for copper. No thermal-diffusion correction was applied, and the deposited energy density obtained from Geant4 was used directly for the temperature-rise calculation. This approach reproduces the assumptions of the original ILC analysis and isolates the accuracy of the Geant4 energy-deposition model from additional effects associated with thermal transport.

The resulting peak temperature rises were found to be in good quantitative agreement with the reference values. For the aluminum spoiler, a maximum temperature rise of $184.6 \pm 4.9^\mathrm{stat.}~\mathrm{K}$ was obtained, compared with approximately $\SI{200}{K}$ reported in Ref.~\cite{FernandezHernando2006}. For the copper spoiler, the simulation yielded $1225.5 \pm 10.7^\mathrm{stat.}~\mathrm{K}$, compared with approximately $\SI{1200}{K}$ reported in the reference calculation. The level of agreement obtained confirms the accuracy of the energy-deposition modeling and provides confidence in the simulation chain used for the ESR collimator survivability studies.

During the benchmarking study, a strong dependence of the peak deposited energy density on the scoring voxel size was observed. Since electromagnetic showers produced by high-energy electrons are concentrated within a narrow beam core, large scoring voxels artificially average the deposited energy over an excessive volume, resulting in an underestimation of the peak energy density and the corresponding temperature rise. The study shows that reliable peak temperature-rise calculations require scoring voxels significantly smaller than the transverse beam dimensions. For the benchmark configurations considered here, convergence of the peak temperature rise was achieved when the voxel dimensions were approximately one-tenth of the incident beam size or smaller. Larger scoring volumes were found to underestimate the peak temperature rise due to spatial averaging of the energy deposition. Consequently, all ESR survivability calculations presented in this work employ scoring voxels satisfying this criterion to ensure accurate estimates of peak energy deposition density and temperature rise.

Figure~\ref{fig:ILC_Benchmark} summarizes the ILC-related benchmark results. In addition to demonstrating good agreement with the published ILC studies, the benchmark establishes the voxel-size requirements used throughout the remainder of this work and provides validation of the simulation chain employed for the ESR collimator survivability analysis.

Having established agreement with the published ILC results, the survivability studies presented in the following sections employ the more detailed enthalpy-based approach described in Section~\ref{sec:TemperatureRiseCalculation}, which accounts for the temperature dependence of the specific heat capacity and phase-transition energies.

\subsection{\label{sec:APS_U_benchmark}Validation Against APS-U Beam-Impact Studies}

While the ILC spoiler benchmark described above validates the accuracy of the Geant4 energy-deposition calculations for a high-energy electron beam interacting with a thin spoiler, additional validation against both independent simulations and experimental observations is required to assess the realism of the temperature-rise and material-survivability methodology employed in this work. For this purpose, the simulation framework was benchmarked against studies performed for the Advanced Photon Source Upgrade (APS-U), which combine PELEGANT beam-dynamics tracking~\cite{Borland2000,Wang2007}, MARS particle-transport simulations~\cite{osti_909902}, and dedicated beam-impact experiments on aluminum absorbers. These studies are particularly relevant to the ESR collimator application because they involve high-energy electron beams, small beam sizes, curved collimator geometries, microsecond-scale beam-loss durations, and damage mechanisms dominated by electromagnetic shower development. The APS-U benchmark therefore provides validation of the complete analysis chain, from beam-loss generation and energy deposition to thermal response and material damage.

\subsubsection{Benchmark Against Static Simulations}

As an initial validation step, the Geant4 framework was used to reproduce the static collimator dose simulations reported in Ref.~\cite{Dooling2022}. The benchmark geometry was implemented following the MARS model shown in Fig.~2 of Ref.~\cite{Dooling2022}. The simulated collimator consisted of pure aluminum with a radius of curvature of \SI{0.865}{m}. A \SI{6}{GeV} electron beam with transverse dimensions $\sigma_x=\SI{13.3}{\micro\meter}$ and $\sigma_y=\SI{11.2}{\micro\meter}$ impacted the collimator at a depth of \SI{0.6}{mm} relative to the apex. The beam intensity corresponded to a single APS-U bunch containing \SI{15.3}{nC}, equivalent to approximately \SI{4.2}{mA} in a 48-bunch fill pattern with a revolution period of \SI{3.68}{\micro s}. Consistent with the original study, temperature-rise calculations employed temperature-dependent aluminum specific heat capacity data, and no thermal-diffusion correction was applied. This preserves the assumptions of the original APS-U analysis and enables a direct comparison between the Geant4 and MARS predictions of dose deposition and temperature rise.

\begin{figure}[htbp]
\centering
\includegraphics[width=\linewidth]{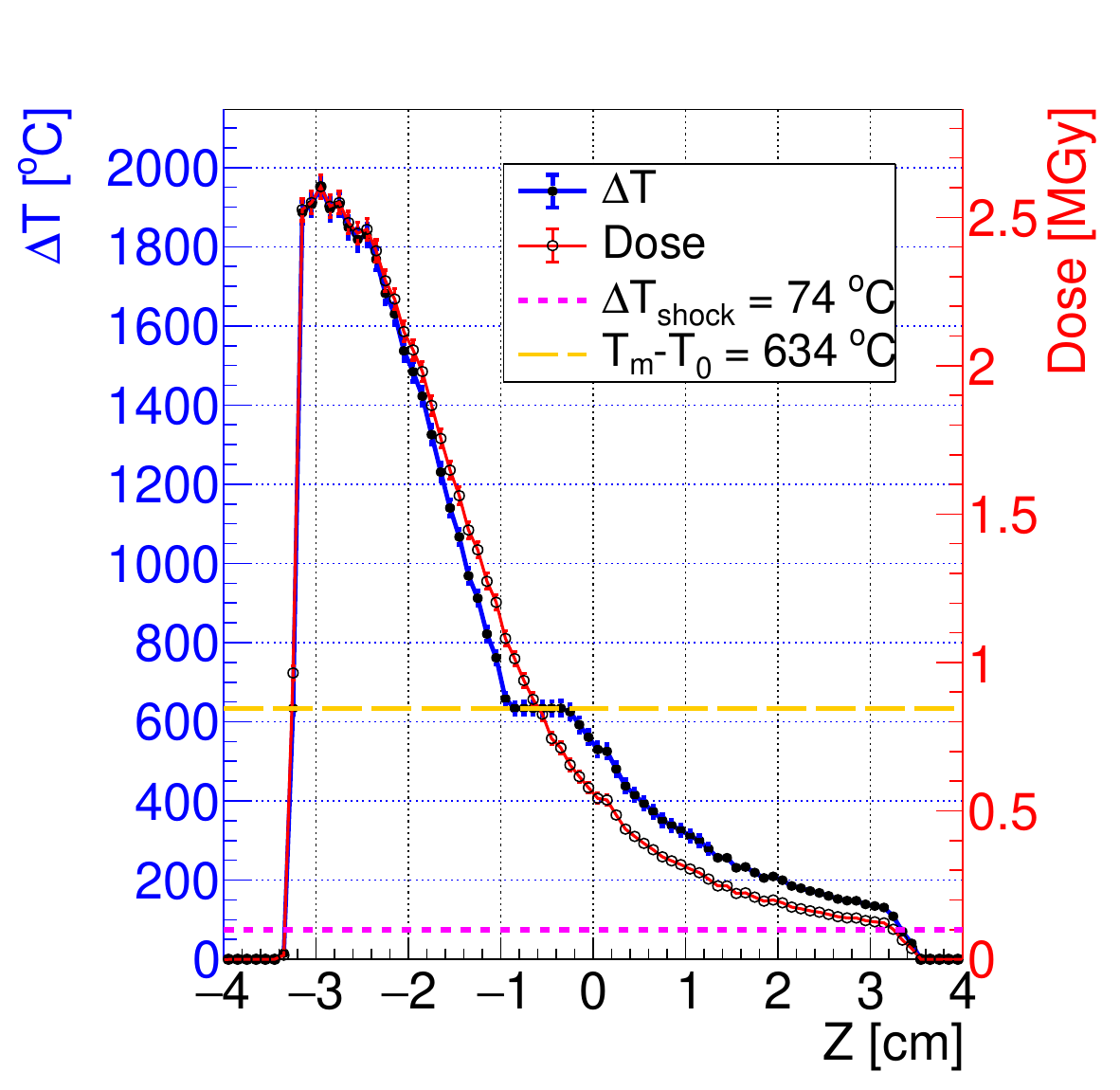}
\caption{\label{fig:Temp_APSU_Al}
Geant4 simulation results of the temperature-rise and deposited-dose distributions along the aluminum spoiler for a single \SI{15.3}{nC} (\SI{4.2}{mA}) \SI{6}{GeV} electron bunch, reproducing the APS-U static MARS benchmark configuration. Longitudinal profiles were obtained using high-resolution scoring voxels of $1 \times 1 \times 1000~\mathrm{\upmu m}^3$ and $10^7$~primary electrons, enabling a detailed comparison with the published predictions of peak dose and temperature rise.}
\end{figure}

Figure~\ref{fig:Temp_APSU_Al} shows the longitudinal distributions of deposited dose and temperature rise obtained with the present Geant4 model. The calculated peak temperature rise was $\Delta T_{\rm max}=1952\pm29$~K, corresponding to a peak material temperature of approximately \SI{2250}{K} for an initial temperature of \SI{300}{K}. This result is in excellent agreement with the published MARS prediction of $T_2=\SI{2285}{K}$. The maximum deposited dose is $D_{\rm max}=2.60\pm0.04$~MGy. This value lies between the peak dose of approximately \SI{1.9}{MGy} reported in the initial APS-U studies and the revised value of approximately \SI{3.1}{MGy} obtained in subsequent PELEGANT-MARS simulations using updated beam-loss distributions. Despite these differences in the reported reference values, the longitudinal distributions of deposited dose and temperature rise are reproduced well, indicating consistent modeling of both the magnitude and spatial development of the electromagnetic shower.

The benchmark confirms that the Geant4 framework accurately reproduces the energy-deposition and thermal-response predictions obtained with MARS for high-energy electron impacts on curved aluminum absorbers. Consistent with the conclusions of the APS-U study, the calculated temperatures substantially exceed the melting temperature of aluminum, indicating that the absorber would not survive even a single-bunch impact under these conditions. The agreement obtained for both the peak temperature and deposited dose provides independent validation of the energy-deposition and temperature-rise methodology employed throughout the ESR collimator studies.

\subsubsection{Benchmark Against Beam-Abort Experiments}

\begin{figure*}[htbp]
\centering
\subfloat[\label{fig:Volm_APSU_Al6061T6}]{\includegraphics[width=0.5\linewidth]{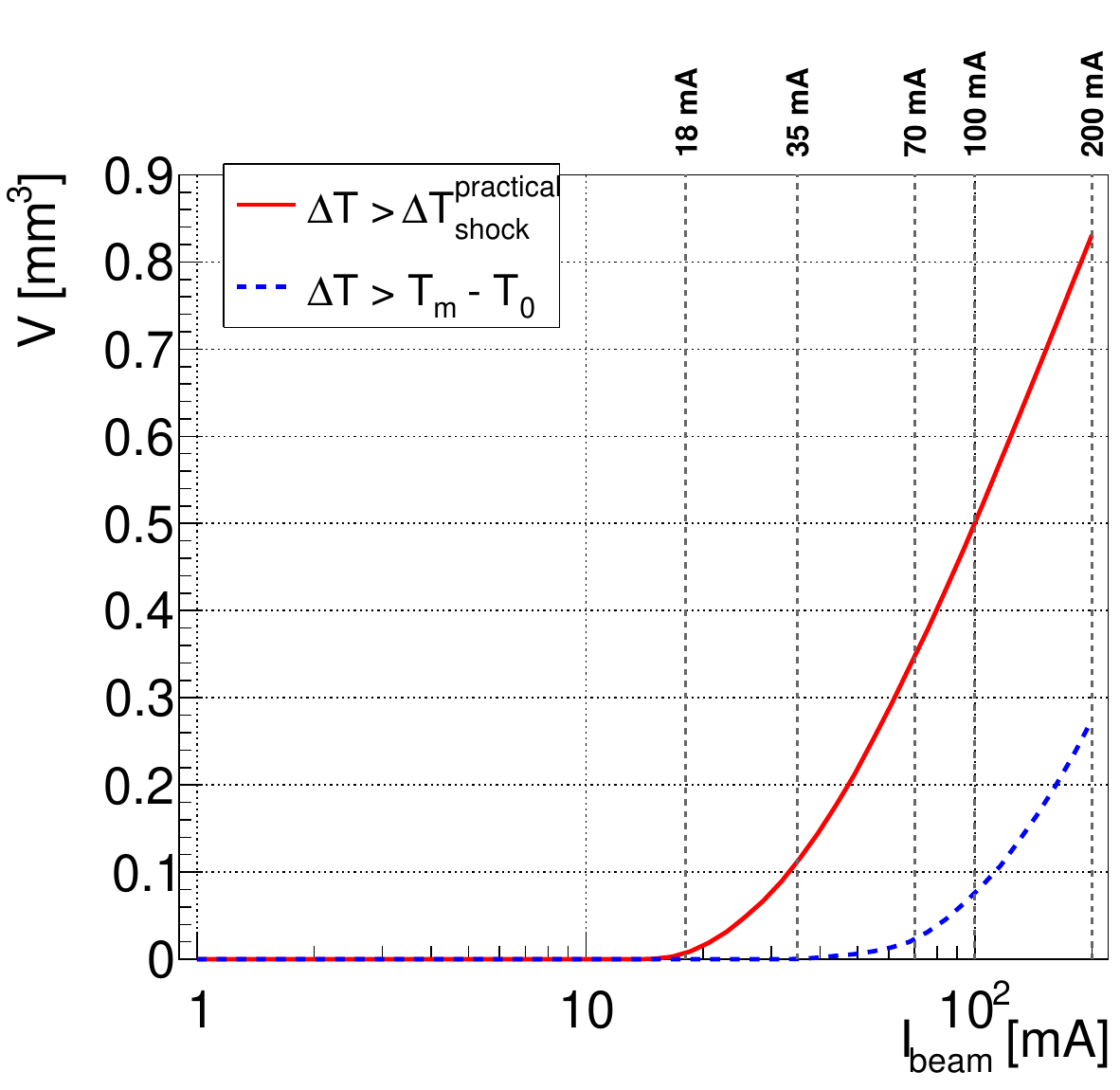}}
\hfill
\subfloat[\label{fig:APSU_EXP_photo}]{\includegraphics[width=0.5\linewidth]{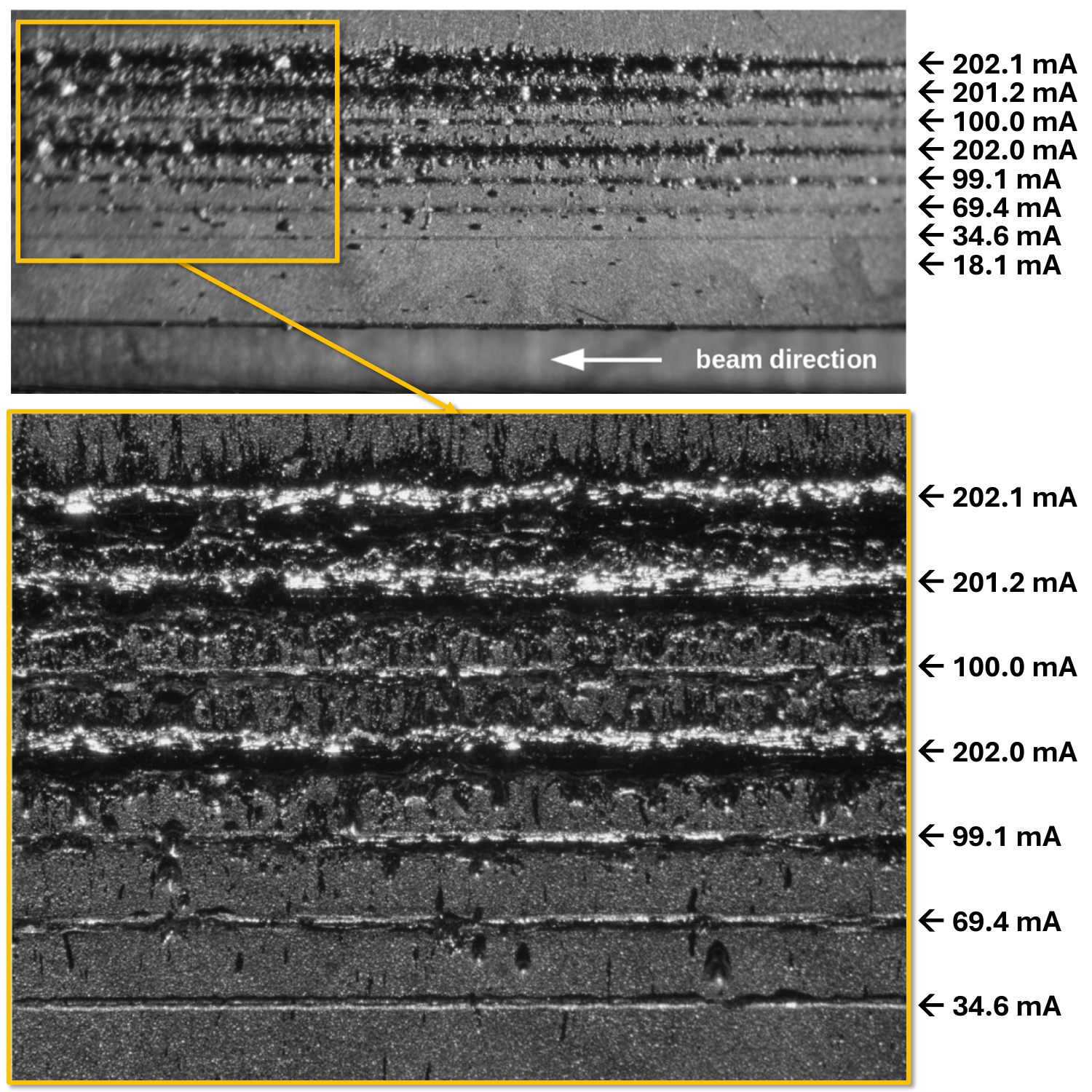}}
\caption{\label{fig:APSExperimentBenchmark}
Comparison of the predicted material response with observations from the APS-U irradiation campaign. (a) Calculated volume of Al6061-T6 exceeding the practical shock limit (red) and melting threshold (blue dashed) as a function of incident beam current. Vertical dashed lines indicate the beam currents investigated during the APS-U irradiation experiments. (b) Photograph of the aluminum collimator sample irradiated during the APS-U beam-abort experiments. The corresponding dumped beam currents are indicated on the right. Adopted from Ref.~\cite{Dooling2022}.
}
\end{figure*}

The second benchmark considered the APS-U collimator irradiation experiments, in which controlled beam aborts were used to investigate beam-induced damage in Al6061-T6 absorbers under realistic storage-ring operating conditions. Unlike the static benchmark discussed above, these studies involve beam losses distributed over several machine revolutions and therefore provide an experimental validation of the thermal-diffusion and survivability methodology developed in this work.

The benchmark geometry reproduced the aluminum collimator test piece used during the APS-U irradiation campaign. The absorber was modeled with a radius of curvature of \SI{0.8}{m} and irradiated by a \SI{6}{GeV} electron beam with transverse dimensions $\sigma_x\approx \SI{110}{\micro\meter}$ and $\sigma_y\approx \SI{8}{\micro\meter}$. The beam impact position was located approximately \SI{0.2}{mm} from the collimator apex. Temperature-rise calculations employed the enthalpy-based methodology described in Section~\ref{sec:TemperatureRiseCalculation}, including temperature-dependent specific heat capacity and latent heat of fusion.

To reproduce the experimentally observed beam-abort dynamics, the beam impact location was sampled from a Gaussian loss-time distribution centered at the third revolution turn following RF muting. A horizontal orbit drift of \SI{100}{\micro\meter} per turn was assumed based on APS measurements, resulting in a continuously varying impact depth during the beam dump. The total beam-loss duration was taken to be approximately five turns ($t_{\mathrm{loss}}=\SI{18.4}{\micro s}$), consistent with the measured APS-U loss profiles.

Because the beam-loss duration is sufficiently long for thermal diffusion to influence the deposited-energy distribution, the APS-inspired diffusion model described in Section~\ref{sec:TemperatureRiseCalculation} was applied. Using the thermal diffusivity of Al6061-T6, $D_{\rm diff} \approx 6.9\times10^{-5}$~m$^2$/s, the deposited-energy was scaled according to the characteristic diffusion length and subsequently processed using the enthalpy-based temperature-rise model. This approach provides a simplified yet computationally efficient approximation of thermal transport during microsecond-scale beam losses.

The APS-U team analyzed the same beam-abort experiments using a sophisticated simulation chain combining PELEGANT beam-dynamics tracking, MARS particle-transport calculations, and FLASH hydrodynamic simulations~\cite{Fryxell2000,Tzeferacos2018}. In contrast, the methodology developed in the present work relies on Monte Carlo energy-deposition calculations, diffusion-based scaling, and enthalpy-based temperature-rise estimates. While significantly simpler and computationally less demanding, the approach captures the principal mechanisms governing beam-induced material damage and, as shown below, reproduces the experimentally observed damage progression with remarkable accuracy.

Figure~\ref{fig:APSExperimentBenchmark} compares the predicted material response with observations from the APS-U irradiation campaign. Figure~\ref{fig:Volm_APSU_Al6061T6} shows the calculated volume of material exceeding the practical shock limit ($\Delta T>\Delta T_{\mathrm{shock}}^{\mathrm{practical}}$) and the volume exceeding the melting threshold ($\Delta T>T_{\mathrm{m}}-T_0$) as a function of incident beam current ($I_\mathrm{beam}$). The predicted thresholds closely follow the experimentally observed damage progression reported in Ref.~\cite{Dooling2022}.

For beam currents below approximately \SI{18}{mA}, the calculated temperature rise remains below the practical shock limit and no visible damage was observed experimentally. At approximately \SI{35}{mA}, a finite volume exceeds the practical shock criterion while remaining below the melting threshold, indicating the onset of the severe thermomechanical loading regime represented by the adopted screening criterion. Consistent with this prediction, only minor surface damage was observed on the irradiated sample. The observed transition between the \SI{18}{mA} and \SI{35}{mA} cases provides experimental support for the order of magnitude and engineering applicability of the practical shock criterion adopted in this work for ductile metallic materials discussed in Section~\ref{sec:TemperatureRiseCalculation}. At approximately \SI{70}{mA}, the calculations predict the onset of melting within a localized region of the absorber, corresponding to the first experimentally observed evidence of melting. For beam currents of \SI{100}{mA} and \SI{200}{mA}, the predicted melted volume increases rapidly, indicating substantial material melting and irreversible damage. This behavior is fully consistent with the extensive melted regions observed on the APS-U test samples.

Although the predicted damaged volume remains below \SI{1}{mm^3}, it is concentrated within a highly localized region surrounding the beam impact point owing to the extremely small APS-U beam dimensions ($\sigma_x\approx\SI{110}{\micro\meter}$ and $\sigma_y\approx\SI{8}{\micro\meter}$). Consequently, even sub-mm$^3$ damaged volumes correspond to significant localized melting and plastic deformation, consistent with the experimentally observed surface damage.

The observed agreement demonstrates that the combined Geant4 energy-deposition model, thermal-diffusion correction, and enthalpy-based survivability methodology provide realistic predictions of beam-induced material damage over a broad range of beam-loss intensities. In particular, the benchmark shows that a comparatively simple framework can successfully reproduce the experimentally observed transition from no visible damage to material softening and ultimately melting without the need for computationally intensive hydrodynamic simulations. The agreement further provides experimental support for the use of a practical shock criterion equal to twice the theoretical thermal-shock limit for ductile metallic materials and alloys. Consequently, the APS-U comparison provides strong validation of the survivability methodology and supports its application to the ESR collimator studies presented in the following sections.

\subsection{\label{sec:SurvivabilityStudy}Survivability Study}

The survivability of the candidate collimator materials was evaluated under abnormal beam-loss conditions representative of accidental beam impacts on the ESR vertical collimator D04V1. The simulations considered a \SI{10}{GeV} electron beam with transverse beam sizes of $\sigma_x=\SI{958}{\micro\meter}$ and $\sigma_y=\SI{340}{\micro\meter}$ at the collimator location. The jaw geometry consisted of a tapered absorber with a taper angle satisfying $\tan(\theta)=1/10$ and a jaw-tip length equal to one radiation length ($X_0$) for each material. Following the convention used in the ILC spoiler survivability study, the incident beam was vertically offset by \SI{2}{mm} relative to the jaw edge. This offset is used as a conservative reference impact condition, ensuring that the beam core is well within the absorber rather than grazing the jaw surface, and therefore represents a severe beam-interception scenario for evaluating material survivability. Temperature-rise calculations employed the benchmarked enthalpy-based methodology described in Section~\ref{sec:TemperatureRiseCalculation}, which was previously validated against published ILC calculations and APS-U beam-impact experiments.

Unless otherwise stated, the reference case corresponds to a beam current of \SI{2}{mA}, approximately equivalent to a single ESR bunch with a charge of \SI{28}{nC}. This single-bunch case is used as a reference unit for quantifying the accumulated beam impact; the beam-current scans presented below correspond to progressively larger numbers of intercepted bunches during an abnormal beam-loss event. Since the bunch charge is fixed, the incident beam current is directly proportional to the number of bunches intercepted by the collimator and is therefore used throughout this section as a convenient measure of beam-impact intensity.

Unlike the APS-U beam-abort experiments discussed in the previous section, ESR accident scenarios are expected to develop within approximately one machine revolution ($t_{\rm loss}\approx\SI{12.8}{\micro s}$), leaving little time for thermal diffusion to redistribute the deposited energy. For the candidate materials considered here, the characteristic thermal diffusion length during one revolution is approximately \SI{30}{\micro\meter}, substantially smaller than the transverse beam dimensions ($\sim\SI{0.9}{mm}\times\SI{0.3}{mm}$). Consequently, the thermal-diffusion correction introduced in Section~\ref{sec:TemperatureRiseCalculation} modifies the peak deposited-energy density by less than approximately 3\%, corresponding to a diffusion correction factor of $f\approx0.97$-0.99 depending on the material. Although the thermal-diffusion correction is included throughout the calculations for completeness, its influence on the predicted temperature rise and survivability limits is negligible under the fast beam-loss scenarios considered for the ESR.

\begin{figure*}[htbp]
\centering
\includegraphics[width=\linewidth]{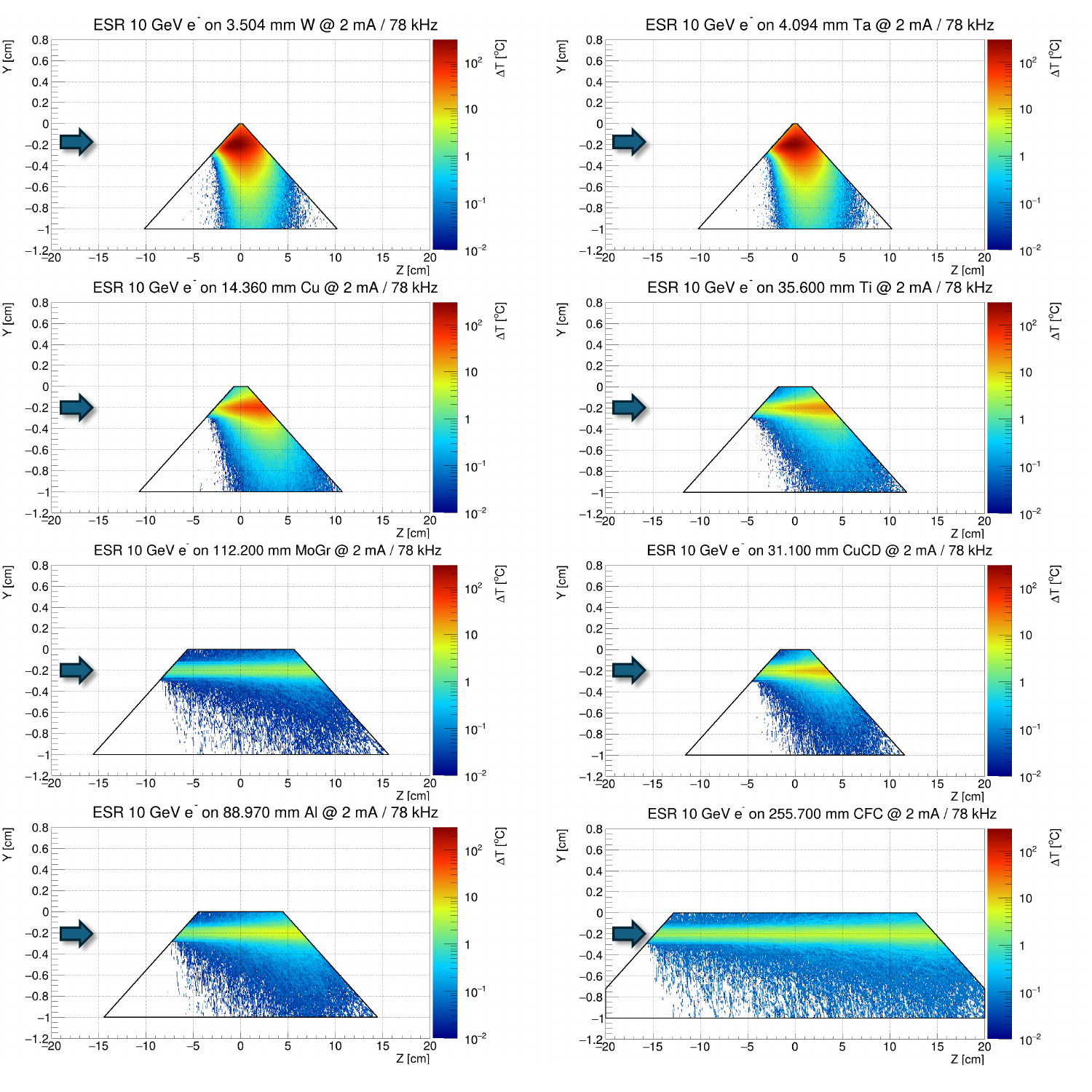}
\caption{\label{fig:ESRTempRiseMap}
Two-dimensional temperature-rise distributions in the jaw tip of the candidate ESR collimator materials following a \SI{2}{mA} beam impact at \SI{10}{GeV}. The beam propagates from left to right.
}
\end{figure*}

Figure~\ref{fig:ESRTempRiseMap} presents the simulated temperature-rise distributions for the candidate absorber materials. As expected, the deposited-energy profile depends strongly on the material radiation length and density. High-$Z$ materials such as tungsten and tantalum concentrate the electromagnetic shower within a relatively small volume, producing localized temperature rises approaching \SI{200}{\celsius} for the reference beam impact. In contrast, graphite-based materials distribute the deposited energy over a substantially larger volume, resulting in peak temperature rises below approximately \SI{10}{\celsius}. Aluminum, titanium, and copper exhibit intermediate behavior, combining significantly lower peak temperature rise than high-$Z$ materials with absorber lengths that remain compatible with the available longitudinal space in the ESR collimator chamber.

\begin{figure*}[htbp]
\centering
\subfloat[\label{fig:ShockVolume}]{\includegraphics[width=0.5\linewidth]{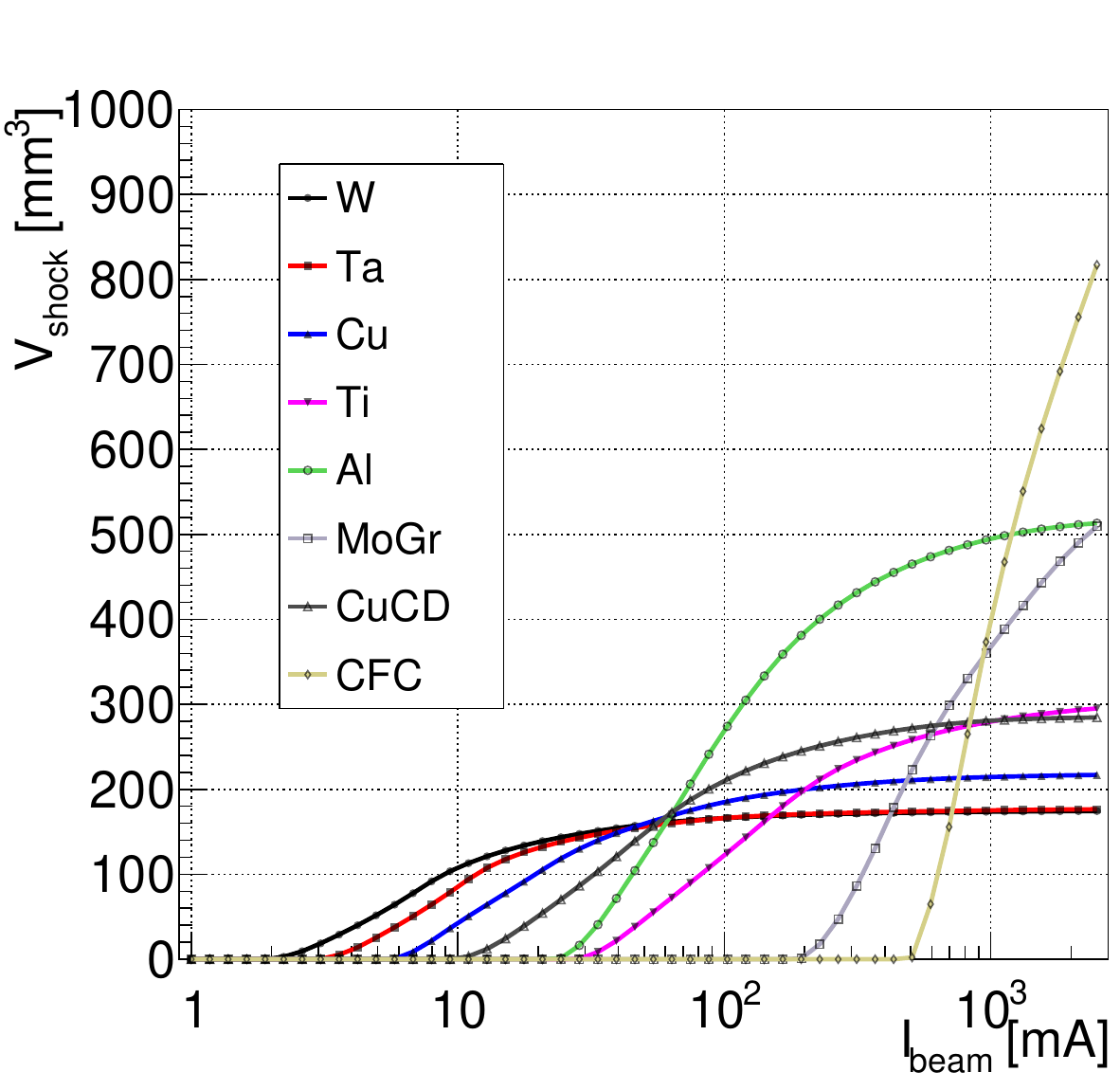}}
\hfill
\subfloat[\label{fig:MeltVolume}]{\includegraphics[width=0.5\linewidth]{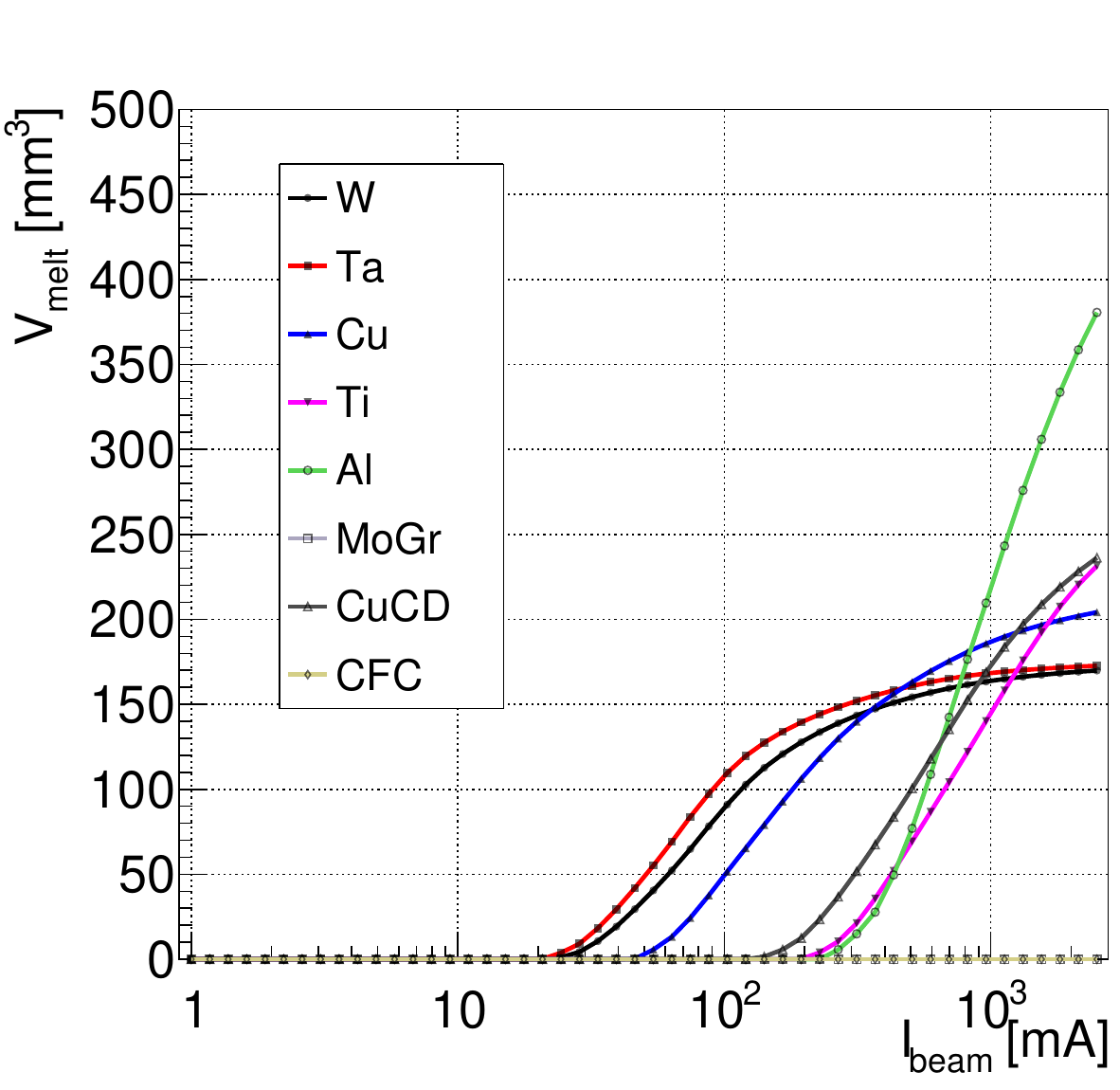}}
\\
\subfloat[\label{fig:ShockVolume_alloys}]{\includegraphics[width=0.5\linewidth]{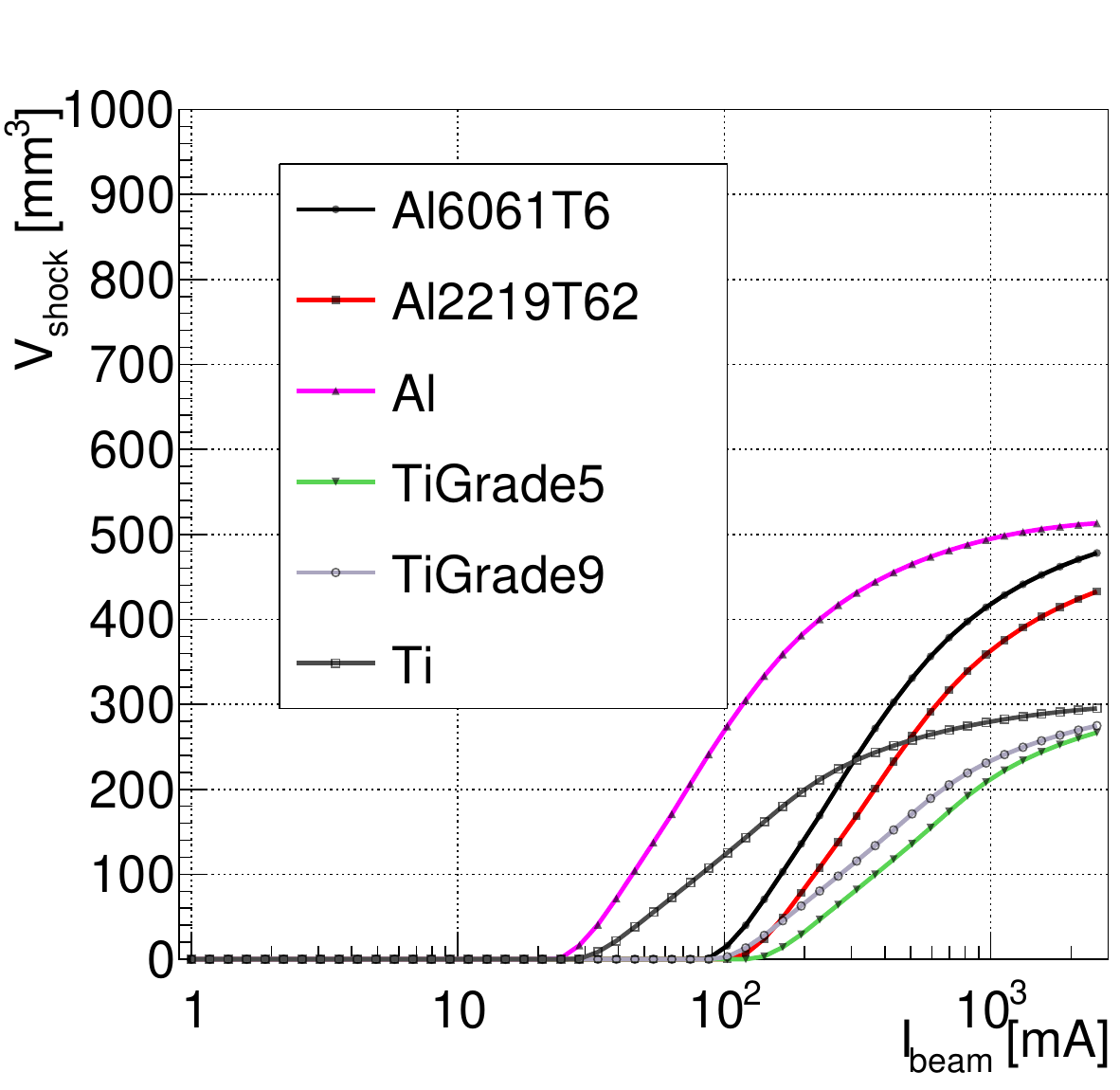}}
\hfill
\subfloat[\label{fig:MeltVolume_alloys}]{\includegraphics[width=0.5\linewidth]{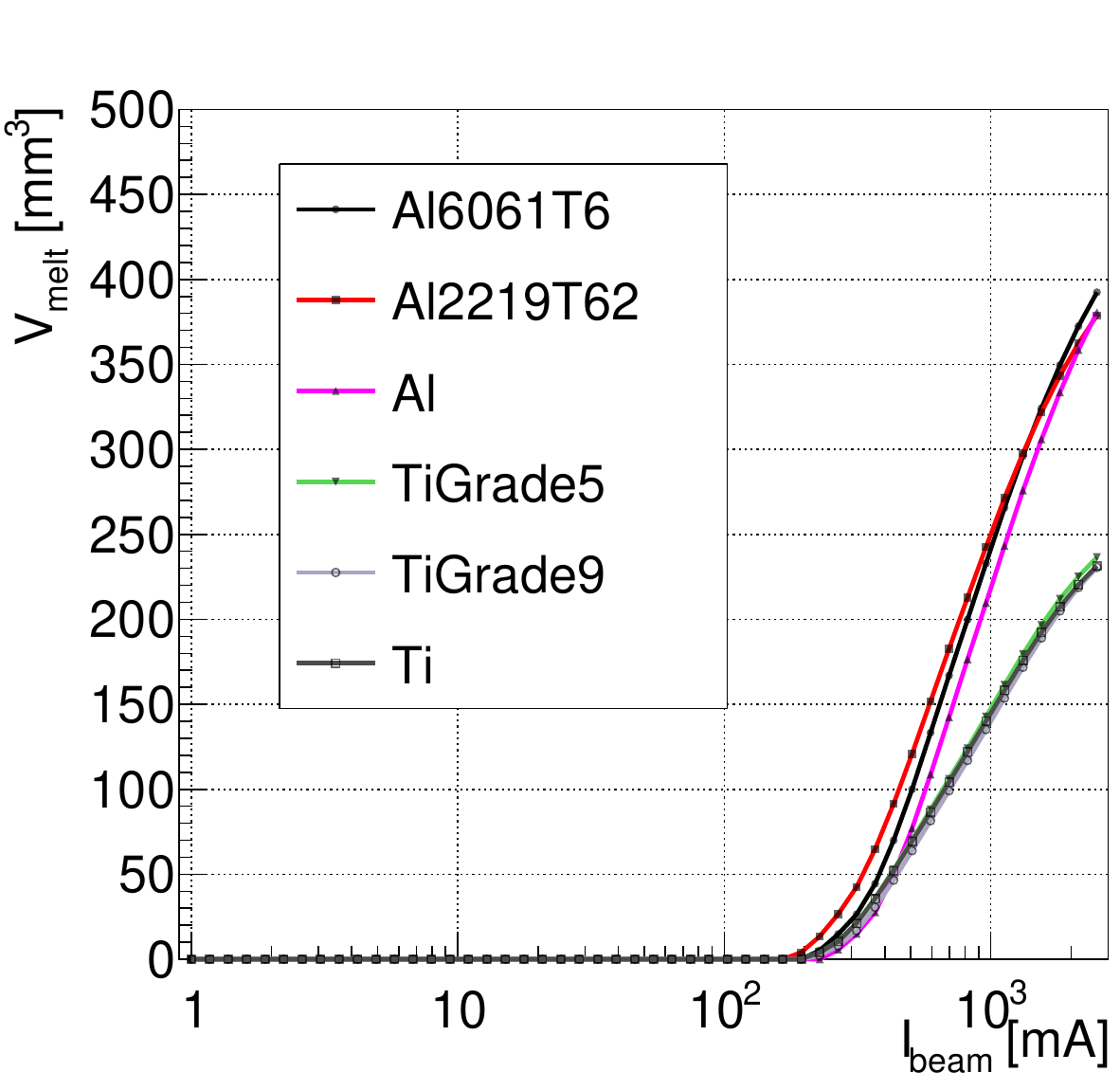}}
\caption{\label{fig:MaterialSurvivability}
Material survivability as a function of incident beam current. (a) Volume exceeding the practical shock limit for the complete material set. (b) Volume exceeding the melting temperature for metallic materials or the sublimation temperature for graphite-based materials. (c) Expanded comparison of aluminum- and titanium-based alloys showing the volume above the practical shock limit. (d) Expanded comparison of aluminum- and titanium-based alloys showing the volume above the melting threshold.
}
\end{figure*}

To quantify survivability within the adopted screening framework, the volume of material exceeding the practical shock threshold and the melting (or sublimation) temperature was evaluated as a function of incident beam current. These quantities provide complementary indicators of the severity of the predicted beam impact: the first identifies regions entering the adopted severe thermomechanical loading regime, while the second identifies regions undergoing an irreversible phase transition. The corresponding material properties relevant to the survivability analysis are summarized in Table~\ref{tab:MaterialProperties}, while Fig.~\ref{fig:MaterialSurvivability} compares the predicted volumes exceeding the respective survivability thresholds for all candidate materials.

High-$Z$ materials such as tungsten and tantalum provide excellent attenuation efficiency because of their short radiation lengths, enabling compact absorber geometries. However, their high density also concentrates the deposited energy within a small volume, resulting in large temperature gradients and severe thermomechanical loading during fast beam impacts. As shown in Figs.~\ref{fig:ShockVolume} and~\ref{fig:MeltVolume}, the predicted practical shock threshold for tungsten and tantalum is reached at beam currents below approximately \SI{10}{mA}, corresponding to fewer than 5 intercepted ESR bunches, while the melting threshold is reached at beam currents below approximately \SI{30}{mA}, corresponding to fewer than 15 bunches.

Operational experience at existing accelerator facilities illustrates these limitations. At SuperKEKB, abnormal beam losses resulted in localized damage to tungsten- and tantalum-based collimator jaws~\cite{PhysRevAccelBeams.27.081001}. Likewise, the APS beam-impact experiments conducted in 2012 demonstrated the damage in tungsten absorbers subjected to the whole-beam abort~\cite{Dooling2022}. Although high-$Z$ materials provide excellent cleaning efficiency, their limited tolerance to accidental beam impacts makes them less suitable for the ESR collimators.

Copper, although formally an intermediate-$Z$ element, behaves similarly to the high-$Z$ materials in this comparison: its density gives it a short radiation length (Table~\ref{tab:MaterialProperties}) that concentrates deposited energy in a small volume, and Figs.~\ref{fig:ShockVolume} and~\ref{fig:MeltVolume} show its shock and melting thresholds reached at beam currents comparable to those of tungsten and tantalum, despite copper's markedly higher thermal and electrical conductivity. Copper is therefore grouped with the high-$Z$ materials for the purposes of this survivability comparison and is not considered further as a baseline absorber candidate.

At the opposite extreme, graphite-based materials such as CFC and MoGr exhibit outstanding beam-impact survivability owing to their low deposited-energy density, high specific heat capacity, and excellent thermal-shock resistance. Even at the highest beam currents considered in this study, the predicted temperatures remain well below the sublimation temperature of either material. Because graphite-based materials do not undergo melting, the dominant failure mechanisms are expected to involve matrix cracking, fiber-matrix debonding, delamination, and progressive degradation of the mechanical properties. The onset of such damage is predicted to occur for beam currents of approximately \SIrange{200}{400}{mA}, corresponding to roughly 100-200 ESR bunches.

Despite their superior survivability, graphite-based materials present several practical challenges for the ESR collimation system. Their large radiation length ($X_0\approx\SI{25}{cm}$) requires substantially longer absorber tips to achieve the desired cleaning efficiency, increasing the longitudinal space required inside the vacuum chamber and tightening alignment tolerances. The increased jaw length also makes the system more sensitive to jaw alignment and beam-trajectory errors. In addition, the relatively low electrical conductivity of both CFC and MoGr necessitates conductive surface coatings to reduce resistive-wall impedance and beam-induced heating. These considerations substantially increase the complexity of the mechanical design and integration.

Intermediate-$Z$ engineering alloys provide a more balanced compromise between collimation performance, beam-impact survivability, vacuum compatibility, mechanical robustness, and impedance. Figures~\ref{fig:ShockVolume_alloys} and ~\ref{fig:MeltVolume_alloys} compare the survivability of the candidate aluminum- and titanium-based alloys. Although titanium alloys exhibit survivability comparable to that of aluminum alloys, their thermal conductivity is approximately an order of magnitude lower, leading to larger local temperature gradients and higher thermally induced stresses during beam impact. The higher thermal conductivity of aluminum alloys promotes a more uniform redistribution of the deposited energy, thereby reducing local thermal gradients and associated thermomechanical stresses. Furthermore, the electrical conductivity of aluminum alloys is approximately an order of magnitude higher than that of titanium alloys, resulting in significantly lower resistive-wall impedance and reduced beam-induced heating. These properties are particularly advantageous for movable collimator jaws operating at small beam apertures.

Among the metallic materials investigated, Al2219-T62 provides the most favorable overall balance of properties. In addition to its excellent thermal and electrical conductivity, the alloy contains less than approximately 1 wt.\% zinc, satisfying the zinc-content requirements commonly adopted for ultra-high-vacuum accelerator components. Aluminum alloys containing larger concentrations of zinc are generally avoided in UHV applications because zinc exhibits a relatively high vapor pressure and may contribute to vacuum degradation during bake-out or beam-induced heating. Al2219-T62 avoids this limitation while maintaining excellent mechanical strength and thermal performance.

The simulations indicate that, under the reference impact conditions considered in this study, Al2219-T62 remains below the adopted practical thermal-shock threshold for beam impacts up to approximately \SI{100}{mA} and below the melting threshold up to approximately \SI{200}{mA}. These beam currents correspond to approximately 50 and 100 ESR bunches, respectively, for a \SI{10}{GeV} electron beam with a bunch charge of \SI{28}{nC}. The first value should be interpreted as a model-dependent screening threshold for severe thermomechanical loading rather than as the onset of yielding or a uniquely defined failure point, whereas the second corresponds to the onset of localized melting within the temperature-rise model.

The practical shock criterion adopted in this work, together with the APS-U benchmark comparison, provides an engineering basis for estimating the available response margin during accidental beam impacts. The corresponding threshold is therefore used as a reference input to the ESR MPS requirements, with the protection system designed to limit the accumulated beam impact below the predicted severe thermomechanical loading regime.

In addition to its favorable survivability, Al2219-T62 permits a practical absorber-tip length of approximately one radiation length (\SI{\sim8}{cm}), providing the required shower development while maintaining a compact jaw geometry. Consequently, the baseline ESR collimator design consists of an Al2219-T62 jaw body with integrated water cooling, synchrotron-radiation absorbers upstream of the horizontal collimators, and a replaceable Al2219-T62 absorber head incorporating an effective shielding length of approximately one radiation length and a 1:10 tapered transition to minimize beam impedance.

Taking into account collimation efficiency, beam-impact survivability, thermal and electrical conductivity, ultra-high-vacuum compatibility, manufacturability, and impedance considerations, Al2219-T62 was selected as the baseline absorber material for the initial ESR collimator implementation. The survivability studies indicate that Al2219-T62 can tolerate beam impacts corresponding to tens to approximately one hundred intercepted bunches at \SI{28}{nC}, covering the range of partial-beam-loss events expected from fast beam instabilities and individual machine-element failures. A full-beam impact at \SI{2.5}{A} represents a substantially more severe and considerably less probable failure scenario and exceeds the survivability limit of the Al2219-T62. The collimator jaw heads are therefore designed to be replaceable, allowing damaged heads to be replaced and providing a straightforward upgrade path to graphite-based materials for operation at higher beam intensities. The proposed mechanical design preserves the possibility of replacing the Al2219-T62 heads with CFC- or MoGr-based components without modifications to the surrounding vacuum chamber or jaw-motion system, allowing the collimation system to evolve as operational experience is gained and the EIC approaches its ultimate beam current and stored-energy goals.

A minimal, self-contained implementation of the temperature-rise and material-survivability workflow used in this study is provided in Ref.~\cite{collimatorTemperatureRise}. The example includes three-dimensional Geant4 energy-deposition scoring for a simplified tapered collimator and ROOT-based~\cite{ROOT_NIMA_1997} conversion of deposited energy density into temperature rise and estimates of the volumes exceeding thermal-shock and melting thresholds. The repository is intended as a reproducibility companion and as a starting point for similar electron-beam collimator studies.

\section{\label{sec:machine_protection}Operational Configuration and Machine Protection}

The survivability studies presented in the previous section establish the beam-impact limits of the candidate absorber materials. The actual beam impact experienced by the ESR collimators, however, depends strongly on the operational jaw settings and on the characteristics of abnormal beam-loss scenarios. Dedicated studies of fast transverse instability-driven beam losses have shown that the original two-collimator configuration does not provide adequate protection under all operating conditions. Detailed analyses of instability growth, beam-loss localization, and machine-protection requirements are presented in Ref.~\cite{osti_3376117}; only the resulting implications for the baseline ESR collimator design are summarized here.

The instability studies identified fast vertical coherent oscillations as the most demanding protection scenario for the ESR. To assess the system under a conservative worst-case condition, an instability growth time of approximately ten turns was considered, under which a significant fraction of the stored beam can be intercepted by the primary collimators within a single revolution, producing beam impacts approaching the survivability limits of metallic absorber materials. The studies further demonstrate that the addition of one horizontal and one vertical collimator, together with optimized jaw settings, effectively localizes beam losses within the dedicated IR4 collimation insertion and reduces loss leakage toward IR6 by more than two orders of magnitude.

\begin{table}[htbp]
    \caption{\label{tab:Collimator_Optics_Parameters_2}
    Updated optics parameters and baseline Day-1 operational settings for the ESR four-collimator configuration at \SI{10}{GeV}, derived from the instability-driven optimization studies reported in Ref.~\cite{osti_3376117}. Here, $S$ denotes the distance from IP6; $\beta$ the betatron function; $\sigma_{\beta} = \sqrt{\beta\varepsilon}$ the RMS transverse beam size due to betatron motion, where $\varepsilon$ is the RMS emittance of 20 and \SI{1.3}{nm} in horizontal and vertical planes, respectively; $D$ the dispersion; and $\mu$ the fractional part of the betatron phase. The subscripts H and V denote the horizontal and vertical planes, respectively. The collimator half-aperture is denoted by $A_\mathrm{coll.}$.}
    \centering
    \begin{tabular}{c|c|c|c|c|c}
        \hline\hline
        Collimator & $S$ & $\beta_\mathrm{H/V}$ & $D_\mathrm{H/V}$ & $\mu_\mathrm{H/V}$ & $A_\mathrm{coll.}$\\ 
        Name & [m] & [m] & [m] & [$2\pi$] & [$\sigma_{\beta}$] ([mm])\\
        \hline
        D04H1 & 569 & 83/35 & 0.1/0 & 0.03/0.67 &  7 (9.1)\\
        D04H2 & 536 & 17/27 & 0.8/0 & 0.30/0.83 & 13 (7.5)\\
        D04V1 & 585 & 46/89 & 0.1/0 & 0.99/0.62 & 19 (6.5)\\
        D04V2 & 517 & 35/41 & 1.1/0 & 0.41/0.93 & 20 (4.6)\\
        \hline\hline
    \end{tabular}
\end{table}

Consequently, the baseline Day-1 ESR collimation system, introduced in Section~\ref{sec:collimation_overview}, consists of four movable collimators: one primary and one additional collimator in each transverse plane. The optimized jaw settings listed in Table~\ref{tab:Collimator_Optics_Parameters_2} maximize cleaning efficiency while maintaining acceptable beam lifetime ($> \SI{5}{hours}$) and operational flexibility, replacing the preliminary two-collimator configuration summarized in Table~\ref{tab:Collimator_Optics_Parameters}.

The instability studies further demonstrate that, although the optimized collimation system effectively localizes beam losses and provides substantial protection of downstream accelerator and detector components, the collimator absorber has a finite impact tolerance under the most severe, rapidly developing beam-loss scenarios. The collimation system must therefore operate as an integral part of the ESR MPS, with the collimators providing the primary beam-loss localization and the MPS limiting the duration and accumulated intensity of abnormal beam impacts.

Dedicated turn-by-turn beam-position monitors (BPMs), installed immediately upstream of each collimator as shown in Fig.~\ref{fig:CollimatorDrawing}, provide predictive detection of coherent orbit excursions before substantial beam interception develops, while fast beam-loss monitors (BLMs) provide an independent confirmation through the observation of precursor losses. The survivability analysis in Section~\ref{sec:SurvivabilityStudy} provides an engineering reference for the magnitude of beam impact that can be tolerated by the absorber under the assumed impact conditions. For the ultimate ESR bunch charge, the practical thermal-shock threshold of the Al2219-T62 absorber corresponds to approximately 50 fully intercepted bunches. This value should not, however, be interpreted as the number of bunches that can be allowed to strike the collimator before initiating a protection action. A rapidly developing beam instability can cause a large fraction of the stored beam to intercept the collimator within a single machine revolution; therefore, the developing beam excursion must be detected before substantial interception occurs.

The studies indicate that the complete protection chain, from detection of the developing beam excursion and generation of the beam-abort request to its transmission and execution by the abort system, should respond within approximately four machine turns~\cite{osti_3376117}. This response time represents a machine-protection design requirement based on the expected turn-scale development of the relevant beam-loss scenarios rather than a direct conversion of the approximately 50-bunch survivability estimate into an allowed response time. The survivability thresholds instead provide an engineering reference for evaluating the available protection margin under accidental beam-impact conditions.

The baseline ESR collimation system therefore combines optimized collimator optics, Al2219-T62 absorber heads, four movable collimators, turn-by-turn BPMs, fast BLMs, and a dedicated beam-abort system into a multilayer machine-protection strategy. This integrated approach provides robust protection of the collimators, SC magnets, vacuum system, and detector infrastructure while preserving a clear upgrade path toward future operation at higher beam currents.

\section{\label{sec:head_refinement}Design refinement of the tapered collimator head}

\begin{figure}[htbp]
\centering
\includegraphics[width=\linewidth]{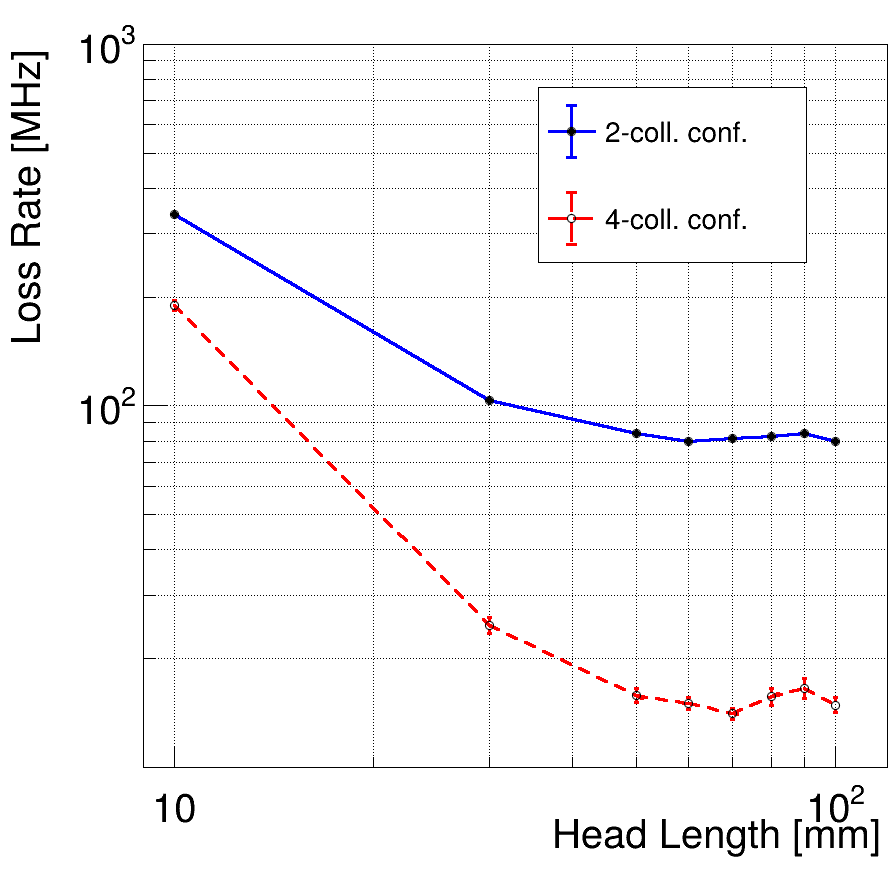}
\caption{\label{fig:IRlossesVsTipLength_10gev_Al2219T62}Total beam loss rate in the IR6 final-focus cryostat region as a function of the Al2219-T62 collimator head length at \SI{10}{GeV}. The blue solid and red dashed curves correspond to two- and four-collimator configurations, respectively.}
\end{figure}

\begin{figure*}[htbp]
\centering
\subfloat[\label{fig:ShockVolume_alloys_Al2219T62}]{\includegraphics[width=0.5\linewidth]{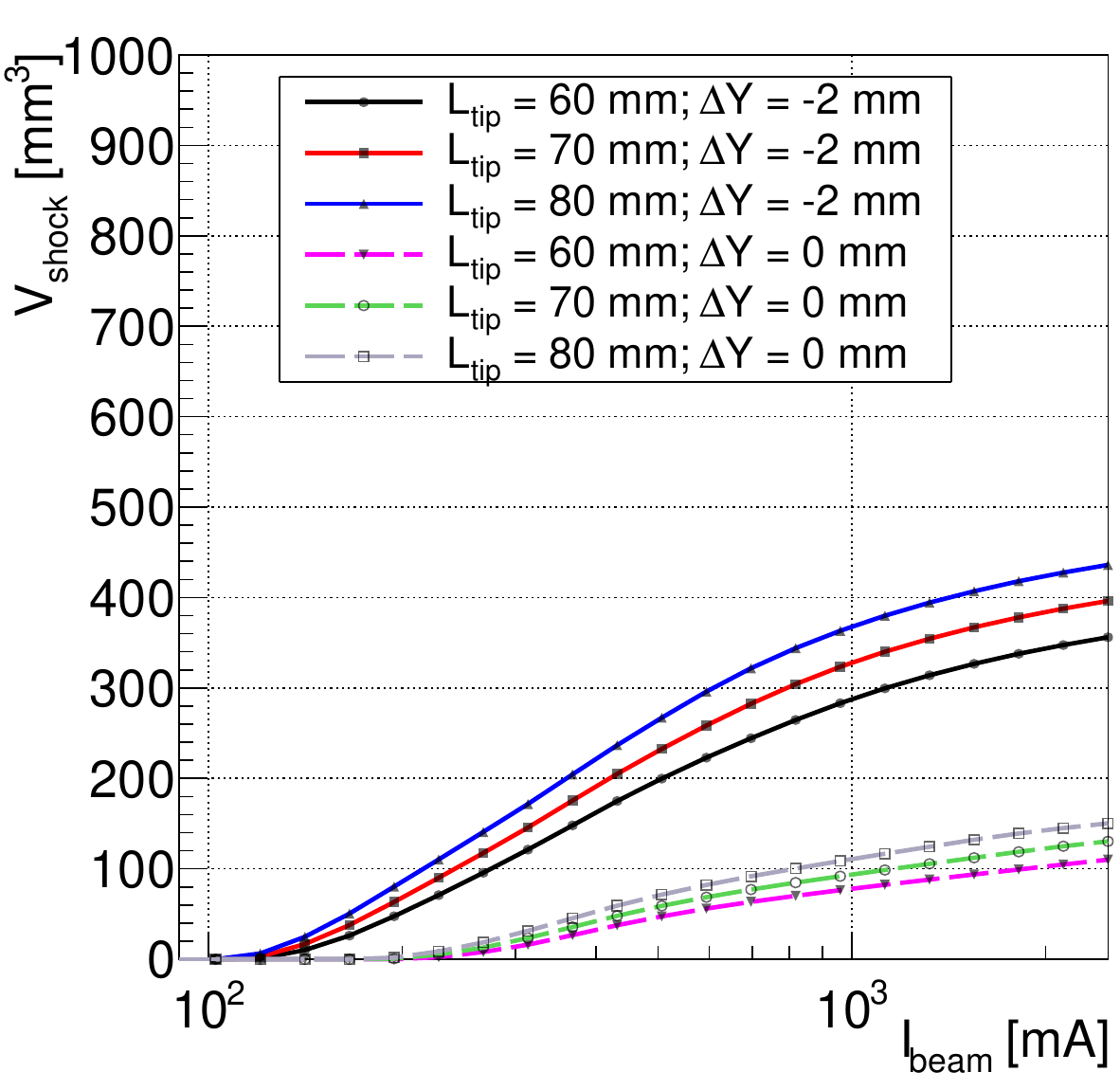}}
\hfill
\subfloat[\label{fig:MeltVolume_alloys_Al2219T62}]{\includegraphics[width=0.5\linewidth]{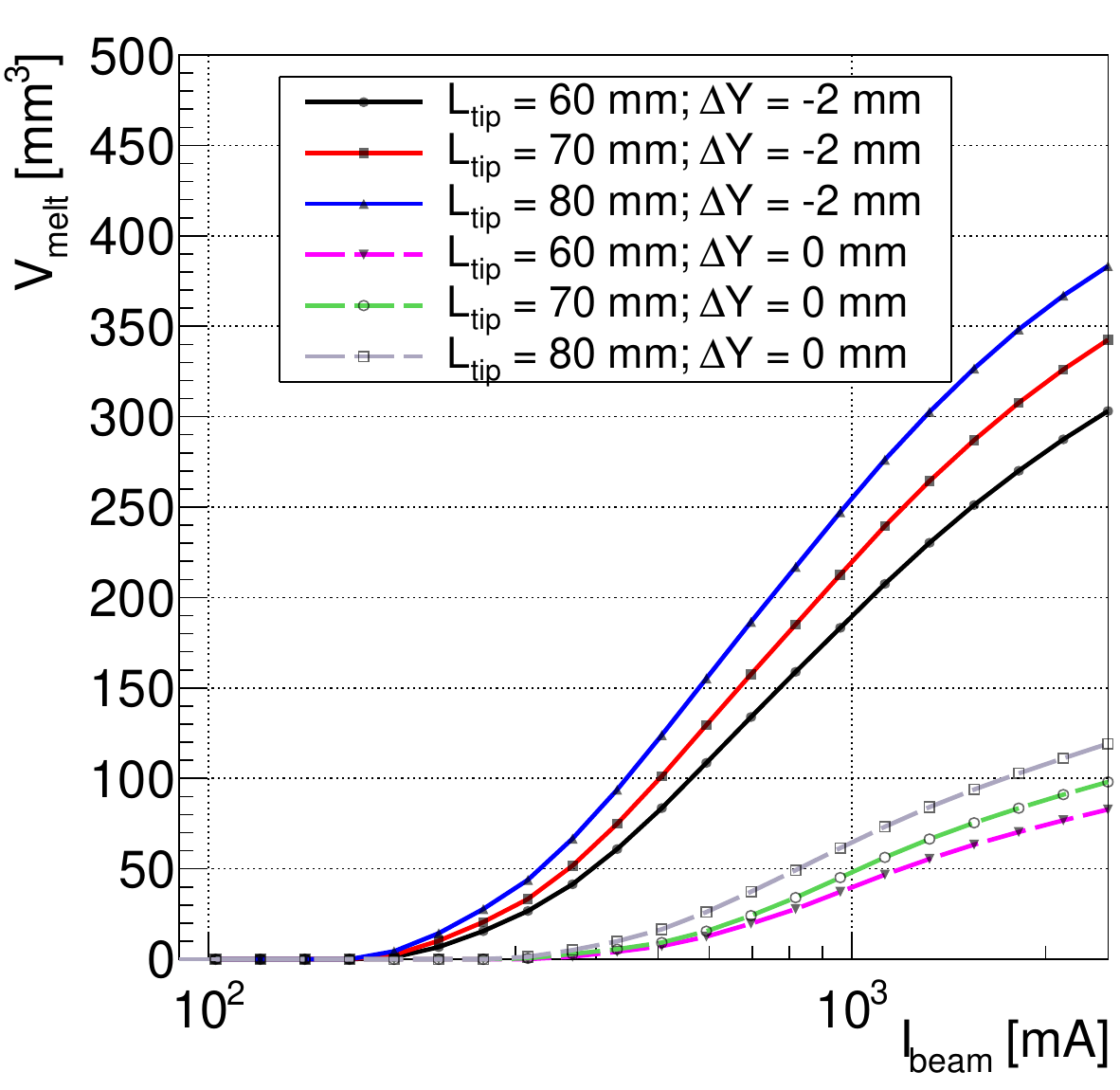}}
\caption{\label{fig:MaterialSurvivability_Al2219T62}
Predicted Al2219-T62 damage volume as a function of the incident beam current and collimator tip length $L_{\rm tip}$ at \SI{10}{GeV} for a bunch charge of \SI{28}{nC}. Results are shown for two vertical beam offsets, $\Delta Y=-\SI{2}{mm}$ (solid curve) and $\Delta Y=\SI{0}{mm}$ (dashed curve), corresponding to complete and approximately half-beam interception by the tapered head, respectively. (a) Volume exceeding the practical shock limit. (b) Volume exceeding the melting temperature.}
\end{figure*}

The baseline ESR collimator design employs a replaceable tapered head with an effective tip length of \SI{80}{mm}, corresponding to approximately one radiation length. This value was originally selected to maximize the attenuation of electromagnetic showers initiated by intercepted beam particles while maintaining acceptable impedance characteristics. The thermomechanical studies presented in the previous sections demonstrate that Al2219-T62 provides substantially greater impact robustness than initially anticipated, while the fast beam-instability studies identify the most demanding beam-impact conditions expected during operation. Together, these results motivated a re-evaluation of the tapered head length to determine whether the baseline design could be further refined without compromising either the collimation performance or the machine protection capability.

To quantify the influence of the tapered head length on collimation performance, additional particle-tracking simulations were performed for head lengths between \SI{10}{mm} and \SI{100}{mm}. The beam-loss distributions were evaluated for both the baseline two-collimator configuration and the upgraded four-collimator configuration using the full design beam intensity under the fast beam-instability conditions described in the previous section, thereby representing a conservative design scenario. The resulting particle loss rates in the SC IR6 cryostat region (from \SI{-15}{\meter} to \SI{22.5}{\meter}), including the ePIC detector, are summarized in Fig.~\ref{fig:IRlossesVsTipLength_10gev_Al2219T62}.

For both collimation configurations, the IR loss rate decreases rapidly as the tapered head length increases from \SI{10}{mm} to approximately \SI{50}{mm}. Beyond approximately \SI{60}{mm}, however, the improvement becomes negligible and the loss rates approach a broad plateau. Increasing the tapered head length above this value produces no significant additional reduction of downstream losses, indicating that additional material beyond approximately \SI{60}{mm} contributes little to further attenuation of the secondary particle shower before particles escape toward the IR.

Based on these results, the subsequent thermomechanical analysis was restricted to tapered head lengths of \SI{60}{mm}, \SI{70}{mm}, and \SI{80}{mm}, for which the IR loss rate had already reached its asymptotic value. Two representative beam-impact scenarios were considered to bound the expected operating conditions. In the first scenario, the beam centroid was vertically offset by $\Delta Y=-\SI{2}{mm}$ with respect to the jaw surface, resulting in complete interception of the beam by the collimator and representing the most conservative impact condition. In the second scenario, the beam centroid was aligned with the jaw edge ($\Delta Y=\SI{0}{mm}$), such that approximately half of the beam impacted the tapered head while the remaining particles propagated downstream. These two cases therefore represent the limiting beam-interception conditions expected during fast beam-loss events.

Figure~\ref{fig:MaterialSurvivability_Al2219T62} compares the predicted molten and mechanically damaged volumes as functions of the beam current for both impact scenarios. Only minor differences are observed among the three head lengths. In particular, the onset of mechanical damage and material melting remains essentially unchanged. Although the shorter tapered heads exhibit slightly smaller damaged volumes at the highest beam currents, the overall thermomechanical response is only weakly dependent on the head length within this range.

The combined collimation and thermomechanical studies therefore indicate that the tapered head length can be reduced from the original \SI{80}{mm} ($\sim1X_0$) to approximately \SI{70}{mm} ($\sim0.8X_0$) without measurable degradation of the collimation performance in either the two- or four-collimator configurations. Since the IR loss rate has already reached its asymptotic value for head lengths exceeding approximately \SI{60}{mm}, further extending the tapered head provides negligible additional shielding benefit. A tapered head length of approximately \SI{70}{mm} therefore represents a favorable engineering compromise, reducing the exposed material volume and geometric impedance while maintaining effective beam-loss mitigation and comparable thermomechanical robustness. The optimization of the tapered head length is particularly important because both the longitudinal length and the electrical conductivity of the jaw material directly influence its resistive wall impedance. As discussed in the following section, these parameters must therefore be considered together when selecting the final tip geometry and material, since unnecessary increases in tip length can introduce an avoidable impedance penalty without providing a corresponding improvement in collimation performance. Based on these results, a tapered head length of \SI{70}{mm} has been adopted as the baseline design for the ESR collimator.

\section{\label{sec:impedance}Impedance Analysis}

The ESR vacuum chamber has a rectangular cross section with horizontal and vertical full apertures of \SI{80}{\milli\meter} and \SI{36}{\milli\meter}, respectively.
The contribution of the ESR collimators to single-bunch beam stability was evaluated in terms of the transverse dipole kick factor, with particular emphasis on the \SI{10}{GeV} operating configuration considered in this work. At \SI{10}{GeV}, the optimized four-collimator configuration listed in Table~\ref{tab:Collimator_Optics_Parameters_2} consists of two horizontal collimators with half-apertures of 9.1 and \SI{7.5}{\milli\meter}, and two vertical collimators with half-apertures of 6.5 and \SI{4.6}{\milli\meter}. The smaller vertical apertures result in stronger wakefields and therefore larger vertical dipole kick factors $k_y$, making the vertical collimators the more important contribution to transverse beam stability. The total beta-weighted vertical kick factor is $\mathcal{K}_y = \sum_j \beta_{y,j} k_{y,j}, $ where the sum extends over the impedance-generating vacuum components, with $k_{y,j}=k_{y,j}^{\mathrm{geom}}+k_{y,j}^{\mathrm{RW}}$ denoting the sum of the geometric and resistive-wall dipole kick factors of component $j$.
In the approximate resistive-wall regime, the transverse contribution scales schematically as 
\begin{equation}
    k_y^{\mathrm{RW}} \propto \frac{L}{b^3}\frac{1}{\sqrt{\sigma_{\rm cond}}},
\end{equation}
where $L$ is the effective conducting length, $b$ is the vertical half-aperture, and $\sigma_{\rm cond}$ is the electrical conductivity~\cite{Chao2023Handbook}. This scaling illustrates the strong sensitivity to aperture, conducting length, and material conductivity.

The wakefields and impedances were calculated using the GdfidL electromagnetic field simulator~\cite{GdfidL}. The longitudinal and vertical dipole pseudo-Green functions were calculated using an rms bunch length of $\bar{\sigma}_s=\SI{0.5}{mm}$, approximately 15 times shorter than the nominal ESR bunch length of $\sigma_s=\SI{7}{mm}$, to resolve the high-frequency wakefield structure relevant to the microwave instability (MWI) and transverse mode-coupling instability (TMCI) analyses.  

In addition to the geometric contribution, the finite electrical
conductivity of the collimator material gives rise to a resistive-wall
impedance. The total longitudinal loss factor can therefore be expressed as
$k_{\rm loss}=k_{\rm loss}^{\rm geom}+k_{\rm loss}^{\rm RW}$,
where $k_{\rm loss}^{\rm geom}$ and $k_{\rm loss}^{\rm RW}$ denote the
geometric and resistive-wall contributions, respectively. For a good conductor in the classical skin-effect regime, the resistive-wall loss factor decreases approximately as 
\begin{equation}
    k_{\rm loss}^{\mathrm{RW}} \propto \frac{L}{b}\frac{1}{\sqrt{\sigma_{\rm cond}}},
\end{equation}
for otherwise identical geometry and bunch parameters. The baseline collimator material, Al2219-T62, has an electrical conductivity of $\sigma_{\rm cond}(\rm Al2219) = (1.74-1.77)\times10^{7}~{\rm S/m}$. The finite-conductivity contribution of Al2219-T62 is also expected to be larger than that of highly conducting materials such as copper. For example, using $\sigma_{\rm cond}(\rm Cu)\approx \SI{5.8e7}{S/m}$ gives 
\begin{equation}
\frac{k_{\rm loss}^{\rm RW}(\mathrm{Al2219})}
     {k_{\rm loss}^{\rm RW}(\mathrm{Cu})}
\simeq
\sqrt{
\frac{\sigma_{\rm cond}(\mathrm{Cu})}
     {\sigma_{\rm cond}(\mathrm{Al2219})}
}
\approx 1.8 .
\end{equation}
for otherwise identical geometry and frequency. The simulated loss factors for the \SI{10}{GeV} beam are approximately \SI{51.9}{mV/pC} and \SI{41.5}{mV/pC} for the vertical collimators with half-apertures of \SI{6.5}{\milli\meter} and \SI{4.6}{\milli\meter}, respectively. These values are comparable to the loss factors reported for SuperKEKB collimators under similar beam parameters in Ref.~\cite{PhysRevAccelBeams.23.053501}. At the nominal single-bunch current of \SI{2.2}{mA}, the bunch length is expected to increase by approximately 10-15\% due to potential-well distortion, indicating a relatively modest impact of longitudinal collective effects at the design operating beam current.

The effective length of the collimator tip, $L_{\rm tip}$, is another important parameter for the impedance. In the resistive-wall regime, the contribution approximately scales with the length of the conducting surface, $k_y \propto L_{\rm tip}$. Therefore, increasing the tip length increases the impedance, and hence the kick factor, approximately linearly, while providing diminishing beam-cleaning benefits once the effective collimation length becomes sufficiently large.

The beam-loss simulations presented in Sec.~\ref{sec:head_refinement} show that the reduction in downstream losses approaches saturation for tip lengths above approximately \SI{60}{mm}. The adopted \SI{70}{mm} tip length therefore provides a compromise between beam-cleaning performance and impedance, while avoiding unnecessary conducting material in the beam environment. Extending the tip to \SI{80}{mm} provides no significant additional collimation benefit while increasing the effective length contributing to the impedance.

The impact of the transverse impedance on single-bunch stability was evaluated in terms of the TMCI. At zero chromaticity, the TMCI threshold can be estimated using the approximate criterion~\cite{Krinsky2005}
\begin{equation}
    \frac{\Delta\nu_y}{\nu_s} = \frac{e I_b}{2E\nu_s\omega_0} \sum_j \beta_{y,j} k_{y,j} \simeq 0.7,
\end{equation}
where $I_b$ is the single-bunch current, $E$ is the beam energy, $\nu_s$ is the synchrotron tune, $\omega_0$ is the revolution angular frequency, and $k_{y,j}$ is the vertical dipole kick factor of the $j$th impedance source. This relation shows that, for a given impedance and beam current, the coherent tune shift increases as the beam energy decreases, resulting in a more restrictive beam-stability condition at \SI{5}{GeV} than at \SI{10}{GeV}.

For the \SI{5}{GeV} ESR configuration, with $\sigma_s=\SI{7}{mm}$, $\nu_s=0.062$, and an average vertical ring beta function of $\langle\beta_y\rangle = \SI{26.7}{m}$, the total beta-weighted vertical kick factor should remain below approximately 
$\mathcal{K}_y^{\rm max} = \SI{96}{kV/pC}$ for the nominal single-bunch current of $I_b=\SI{2.2}{mA}$ in order to keep the TMCI threshold above the design operating current. The present ESR impedance budget corresponds to $\mathcal{K}_y \approx \SI{23.4}{kV/pC}$, yielding an analytical TMCI threshold current of approximately \SI{9}{mA}, well above the nominal operating current.

For the \SI{10}{GeV} configuration, the two vertical collimators, including both geometric and resistive-wall contributions, provide a combined vertical dipole kick factor of approximately $\mathcal{K}_{y,{\rm coll}} \approx \SI{3.5}{kV/pC}$. This contribution represents approximately 15\% of the present beta-weighted vertical impedance budget, indicating that the ESR collimators make a measurable but controlled contribution to the total transverse impedance.

At \SI{5}{GeV}, the vertical collimators are operated with larger half-apertures, as shown in Table~\ref{tab:Collimator_Optics_Parameters}, and are therefore expected to yield a smaller transverse kick factor. We consequently use the kick factor calculated for the more restrictive \SI{10}{GeV} collimator apertures as a conservative estimate of the collimator contribution at \SI{5}{GeV}. This conservative contribution corresponds to only about 3.6\% of the allowable \SI{5}{GeV} kick-factor limit. The vertical collimators therefore remain well within the available impedance margin from the standpoint of transverse single-bunch stability.

The comparison between the two operating energies shows that they present complementary constraints. The smaller apertures at \SI{10}{GeV} make this configuration the more demanding case from the collimator impedance perspective, while the lower beam energy at \SI{5}{GeV} results in a more restrictive overall beam-stability condition. The adopted \SI{10}{GeV} configuration, combining the optimized vertical apertures with the Al2219-T62 absorber material and \SI{70}{mm} tapered-head length, therefore provides the relevant basis for the impedance optimization while maintaining sufficient margin against transverse single-bunch instability.

\section{\label{sec:steady_state}Steady-State Thermomechanical Analysis}

\begin{figure*}[htbp]
\centering
\subfloat[\label{fig:steady_state_temperature}]{\includegraphics[width=\linewidth]{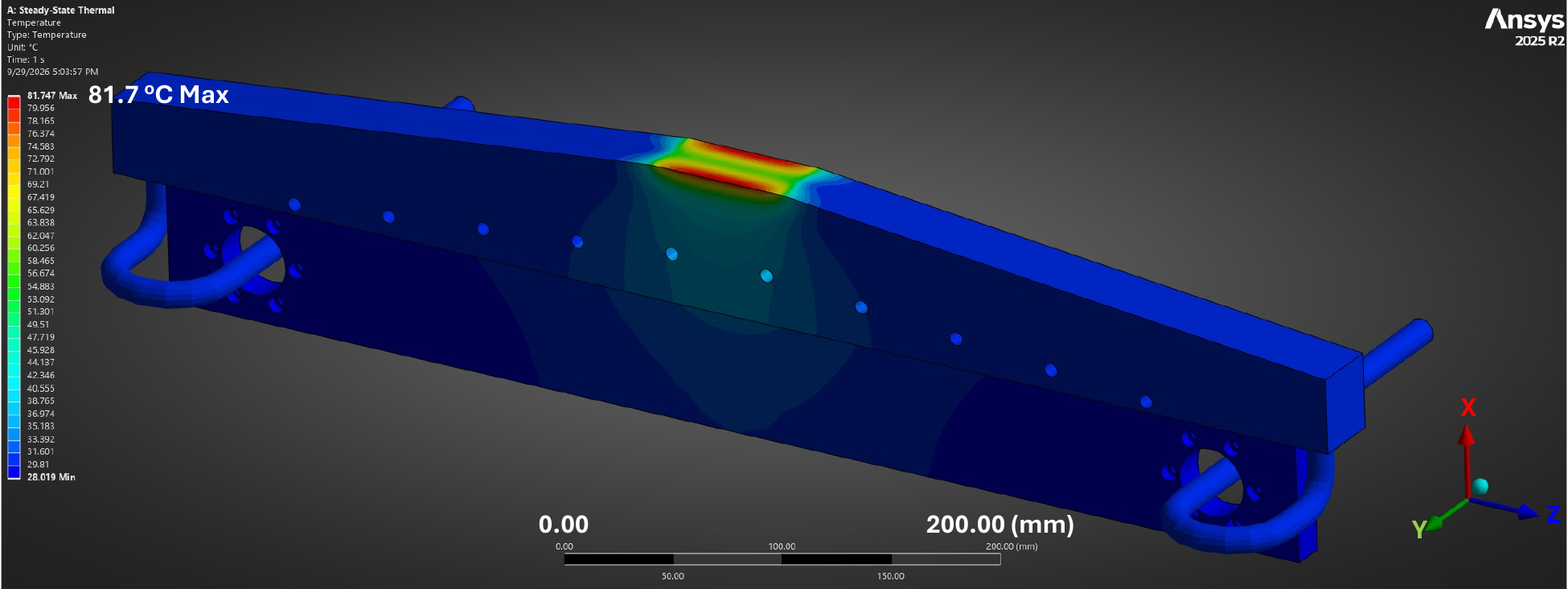}}
\\
\subfloat[\label{fig:steady_state_stress}]{\includegraphics[width=\linewidth]{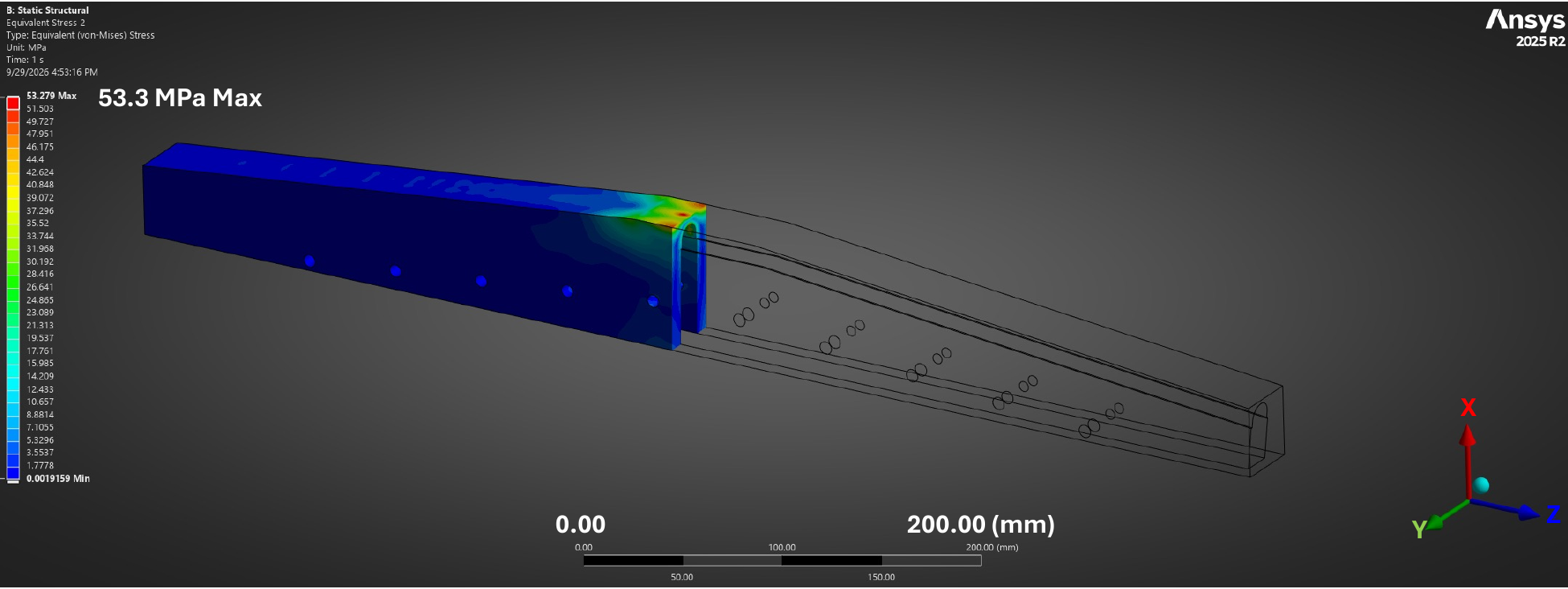}}
\caption{\label{fig:steady_state}
ANSYS simulation results for the steady-state (a) temperature and (b) von Mises stress distributions in the Al2219-T62 ESR collimator under the conservative beam-induced heating scenario.}
\end{figure*}

The thermal performance of the ESR collimator under steady-state conditions was evaluated using ANSYS engineering simulation software~\cite{ANSYS2025R2} for two beam-induced heating scenarios. 

First, the power deposition associated with beam losses was obtained from the tracking simulations described in Ref.~\cite{osti_3021706}, which account for losses due to Touschek scattering, beam-gas interactions, scattering on thermal radiation, and Bethe-Heitler processes at the IP. The losses intercepted by the collimators were evaluated using the aperture settings listed in Table~\ref{tab:Collimator_Optics_Parameters_2}. Although the maximum incident beam power on a collimator is below \SI{20}{W}, only a small fraction of the intercepted beam power is absorbed by the collimator jaws. The resulting heat load on the primary collimator jaws is only $\mathcal{O}(\SI{200}{mW})$, producing a steady-state temperature rise of less than \SI{1}{\degreeCelsius}. 

The much larger heat load associated with beam-induced wakefields was then considered. The total heating power deposited in the collimator assembly due to the beam-induced loss factor is given by~\cite{Wiedemann2015}
\begin{equation}
    P_{\rm loss} = k_{\rm loss}\frac{I^2 T_{\rm rev}}{n_{\rm b}},
\end{equation}
where $I=\SI{2.5}{A}$ is the total beam current, $T_{\rm rev}\approx\SI{12.8}{\micro\second}$ is the revolution period, and $n_{\rm b}=1160$ is the number of bunches. For the maximum loss factor of $k_{\rm loss}=\SI{51.9}{mV/pC}$ discussed in Sec.~\ref{sec:impedance}, the corresponding heating power is $P_{\rm loss}^{\rm max} \approx \SI{3.6}{kW}$ per collimator assembly.

In reality, the beam-induced power is distributed among several components of the collimator assembly, including the beam pipe, tapered and flat jaw surfaces, and bellows. To obtain a conservative estimate of the thermal and mechanical response of the jaw, the entire \SI{3.6}{kW} beam-induced power was assumed to be deposited only on the two jaw tips, with the load distributed equally between the opposing jaws. Thus, a heat load of \SI{1.8}{kW} was applied uniformly to the tip of each jaw over an area of approximately \SI{23}{cm^2}. This assumption represents a conservative upper-bound estimate, since the actual beam-induced power will be distributed over a larger volume and among multiple components of the collimator assembly.

The steady-state thermal analysis included radiative heat transfer from all external collimator jaw surfaces to the surrounding environment, assumed to be at \SI{22}{\degreeCelsius}. Heat removal through the internal cooling channels was modeled using water with an inlet temperature of \SI{30}{\degreeCelsius}, a volumetric flow rate of \SI{2.5}{gal/min}, and a corresponding flow velocity of approximately \SI{2}{\meter/\second}. These boundary conditions are representative of the assumed EIC tunnel environment and technical cooling-water conditions.

Figure~\ref{fig:steady_state_temperature} shows the resulting steady-state temperature distribution obtained from the ANSYS analysis. The maximum temperature is approximately \SI{82}{\degreeCelsius}, with the highest temperature localized near the jaw-tip region where the heat load is applied. The remainder of the jaw remains at substantially lower temperature owing to heat conduction through the jaw body and heat removal through the cooling system.

The corresponding thermally induced mechanical stresses in the Al2219-T62 jaw were evaluated using the temperature distribution obtained from the thermal analysis. Figure~\ref{fig:steady_state_stress} shows the resulting von~Mises stress distribution. The maximum calculated stress is approximately \SI{53}{MPa} and remains below the yield and ultimate tensile strength of \SI{290}{MPa} and \SI{414}{MPa}, respectively, for Al2219-T62. The jaw therefore remains within the elastic regime under the assumed steady-state thermal load.

The analysis demonstrates that the baseline Al2219-T62 jaw can withstand the thermal load associated with the maximum beam-induced loss factor considered in this study without exceeding its yield strength. The baseline collimator incorporates internal water-cooling channels with a cross-sectional area of approximately \SI{0.8}{cm^2}, while cooling-water piping will be provided throughout the EIC tunnel to supply various accelerator systems, including the ESR collimators. Under the conservative heating scenario considered here, the adopted cooling conditions provide sufficient heat removal to maintain the jaw temperature and thermally induced stresses within acceptable limits.

\section{\label{sec:conclusions}Conclusions}

This work presents the design and optimization of the baseline electron-beam collimator for the Electron Storage Ring (ESR) of the Electron-Ion Collider. The design addresses electromagnetic shower containment, absorber survivability under accidental beam impacts, steady-state thermomechanical performance, beam coupling impedance, and compatibility with machine-protection requirements. The resulting analysis establishes a consistent connection between the collimator geometry and material selection, expected beam-loss conditions, and the protection requirements associated with accidental beam impacts.

A Geant4-based methodology was developed to evaluate energy deposition and the resulting instantaneous temperature rise in the collimator jaw head. The approach incorporates three-dimensional energy-deposition scoring, temperature-dependent material properties, phase-transition energies, and thermal diffusion. The simulation methodology was benchmarked against published simulations and beam-impact experiments, providing validation of the electromagnetic shower modeling and temperature-rise calculations. The benchmarking studies also established the spatial-resolution requirements necessary for reliable evaluation of the highly localized peak energy deposition produced by high-energy electron beams. A minimal, self-contained implementation of the temperature-rise calculation workflow was also made publicly available to facilitate the application of this approach to similar studies.

A comparison of candidate absorber materials demonstrated the tradeoff between collimation efficiency and robustness against accidental beam impacts. High-$Z$ materials provide compact electromagnetic-shower containment but experience severe localized heating during beam impacts, whereas graphite-based materials provide greater robustness at the expense of substantially longer absorber lengths and lower electrical conductivity. Among the metallic materials considered, Al2219-T62 provides a favorable overall balance of impact survivability, thermal conductivity, electrical conductivity, mechanical strength, and ultra-high-vacuum compatibility. For the ultimate ESR bunch charge of \SI{28}{nC} at \SI{10}{GeV}, the simulations predict that an Al2219-T62 jaw remains below the adopted practical thermal-shock threshold for approximately 50 fully intercepted bunches and below the localized melting threshold for approximately 100 bunches under the reference impact conditions considered.

The calculated practical thermal-shock threshold provides an important reference input to the machine-protection requirements associated with the collimator. Turn-by-turn beam-position measurements and fast beam-loss monitors can provide predictive and confirmatory detection of developing beam excursions, respectively. The protection strategy is therefore required to limit the accumulated beam impact below the model-dependent severe thermomechanical loading threshold.

The jaw geometry was optimized to balance collimation performance, beam-impact robustness, thermomechanical performance, and beam coupling impedance. The resulting baseline design employs a \SI{70}{mm} tapered Al2219-T62 jaw tip, which provides effective shielding of the interaction region while limiting the exposed material volume and impedance contribution. The impedance analysis shows that the optimized collimator configuration remains compatible with the ESR transverse single-bunch stability requirements. The relatively high electrical conductivity of Al2219-T62, together with the optimized jaw geometry, limits the resistive-wall contribution while preserving sufficient absorber robustness.

The steady-state thermomechanical analysis further demonstrates the robustness of the baseline jaw under beam-induced heating. The power associated with intercepted beam losses produces only a negligible temperature rise in the primary collimator jaws. For the more demanding wakefield-heating scenario, the maximum \SI{3.6}{kW} power calculated for the complete collimator assembly was conservatively assumed to be deposited exclusively on the two jaw tips, with \SI{1.8}{kW} applied uniformly to each jaw. Under the assumed cooling conditions, the maximum calculated temperature is approximately \SI{82}{\degreeCelsius}, and the maximum von Mises stress of approximately \SI{53}{MPa} remains below the material-strength values of Al2219-T62.

The steady-state and accidental beam-impact analyses presented in this work represent complementary conservative scenarios. In the steady-state calculation, the complete beam-induced power of the collimator assembly is conservatively concentrated on the jaw tips. In contrast, the accidental-impact calculations consider a highly localized beam impact over a region of approximately $1\times0.1~\mathrm{mm^2}$, for which the rapid local temperature rise determines the material robustness limit. In the actual collimator, the wakefield power will be distributed among the beam pipe, jaw surfaces, bellows, and other impedance-generating components. Consequently, the local steady-state heat load at the beam-impact location is expected to be substantially smaller than that assumed in the conservative steady-state analysis and to have only a limited influence on the initial material temperature and properties relevant to the localized robustness calculations.

The resulting baseline ESR collimator design combines an Al2219-T62 absorber jaw, an optimized tapered jaw tip, internal water cooling, and material and geometric properties selected to balance beam-loss interception, impact robustness, thermomechanical performance, and beam impedance. The analysis demonstrates that these competing requirements can be addressed within a unified collimator design framework. The methodology and results developed in this work provide the basis for the ESR collimator design and can also be applied to the robustness evaluation and optimization of collimators for other high-current electron storage rings.

\section*{\label{sec:Acknowledgments}Acknowledgments}
We thank our colleagues in the EIC Project for many helpful discussions and for providing technical information on the experimental setup. We also thank the Jefferson Lab Scientific Computing Team for providing access to the JLab computing farm, which enabled the high-statistics, CPU-intensive simulations presented in this work.

We are grateful to S.~Terui and T.~Ishibashi (KEK), S.~Redaelli (CERN), and J.~Dooling (ANL) for valuable discussions, constructive feedback on the methodology, and helpful suggestions throughout this work.

This work was supported by Brookhaven Science Associates, LLC under Contract No.~DE-SC0012704 with the U.S. Department of Energy.

\bibliography{bibfile}

\end{document}